\documentclass[10pt,conference,twocolumn]{IEEEtran}

\usepackage{amsmath}    
\usepackage{amsthm}     
\usepackage{graphicx}   
\usepackage{verbatim}   
\usepackage{hyperref}   
\usepackage{amssymb}

\usepackage{algorithm}
\usepackage{algpseudocode}

\newcommand{\C}{{\mathbb{C}}}  

\newtheorem{lemma}{Lemma}
\newtheorem{claim}{Claim}
\newtheorem{theorem}{Theorem}
\newtheorem{corollary}{Corollary}
\newtheorem{definition}{Definition}
\newtheorem{proposition}{Proposition}

\begin{document}
\title{Interactive Interference Alignment}
\author{\IEEEauthorblockN{Quan Geng, Sreeram Kannan, and Pramod Viswanath}
\IEEEauthorblockA{Coordinated Science Laboratory and Dept. of ECE\\
University of Illinois, Urbana-Champaign, IL 61801\\
Email: \{geng5, kannan1, pramodv\}@illinois.edu}
}

\maketitle

\begin{abstract}
We study interference channels (IFC) where interaction among sources and destinations is enabled, e.g., both sources and destinations can talk to each other using full-duplex radios.  The interaction can come in two ways:  1) {\em In-band interaction:} sources and destinations can transmit and listen in the same channel simultaneously, enabling interaction. 2) {\em out-of-band interaction:} destinations talk back to the sources on an out-of-band channel, possible from white-space channels. The flexibility afforded by interaction among sources and destinations allows for the derivation of interference alignment (IA) strategies that have desirable ``engineering properties'': insensitivity to the rationality or irrationality of channel parameters, small block lengths and finite SNR operations. We show that for several classes of interference channels the interactive interference alignment scheme can achieve the optimal degrees of freedom. In particular, we show the {\em first simple scheme} (having finite block length, for channels having no diversity) for $K=3,4$ that can achieve the optimal degrees of freedom of $\frac{K}{2}$ even after accounting for the cost of interaction. We also give simulation results on the finite SNR performance of interactive alignment under some settings. 

On the technical side, we show using a Gr\"{o}bner basis argument that in a general network  potentially utilizing cooperation and feedback, the optimal degrees of freedom under linear schemes of a fixed block length is the same for channel coefficients with probability $1$. Furthermore, a numerical method to estimate this value is also presented. These tools have potentially wider utility in studying other wireless networks as well.

\end{abstract}

\section{Introduction} \label{sec:intro}
Due to the broadcast nature of wireless communication, an active transmitter will create interference to all other receivers using the same channel  (in time and frequency). This is a basic challenge in  supporting concurrent reliable wireless communications. Traditional methods have involved power control, scheduling and antenna beamforming/nulling to manage interference.  Interference alignment is a novel interference management technique which makes concurrent wireless communications feasible \cite{CJ2008, M-A2007}.

Since its introduction, substantial research has been conducted in understanding the gains of interference alignment. From a theoretical perspective, the focus has been mainly in understanding the degrees of freedom (first term in the high SNR  approximation of the information theoretic capacity region). A seminal result, shown in \cite{CJ2008}, is that we can communicate at  a total $\frac{K}{2}$ degrees of freedom for nearly all  time-varying/frequency selective $K$-user interference channel  (this is also an upper bound). To get to $\frac{K}{2}$ degrees of freedom,  the  scheme employed (vector space interference alignment) requires the channel diversity  to be unbounded; in fact, one needs the channel diversity    to grow like $K^{2K^2}$ \cite{BCT2011}. The diversity order required is huge even for modest number of users $K$ (such as 3 and 4). This is definitely a huge impediment in a practical communication system, where there is hardly enough channel  diversity to make such a vector space interference alignment scheme feasible.

A  practically relevant problem is to understand the  fundamental degrees of freedom for a  fixed deterministic channel.  For fully connected channel matrix $H$, the total  degrees of freedom are upper bounded by $\frac{K}{2}$ and \cite{RealInter2009} shows that this upper bound is achievable for almost all channel matrices $H$ using a coding scheme based on Diophantine approximation. However, this result is limited in two ways. First, the coding scheme is very sensitive to whether entries of $H$ are rational or irrational. Second, although it is provable $\frac{K}{2}$ degrees of freedom are achievable for almost all $H$, for a given $H$, in general it is not known what are the optimal achievable degrees of freedom. \cite{Wu2011} shows that the $\frac{K}{2}$ result for almost all $H$ can be derived using R\'{e}nyi information dimension. Again, the result is sensitive to whether entries of $H$ are rational or irrational, and for fixed channel matrix $H$, in general the optimal achievable degrees of freedom is unknown. The recent works of \cite{NM2012, Or12}   address this issue to a good extent  for the case of the two-user X channel and the symmetric
$K$-user interference channel respectively, but  the engineering implication of the proposed coding schemes remains unclear.

Therefore, despite significant theoretical progress on the $K$-user interference channel problem, it is still unclear how to make interference alignment practical. The drawbacks of existing schemes may be inherent to the  channel model which assumes sources can only transmit and destinations can only listen, while in practice radios can both transmit and receive.
We study new channel models where interaction among sources and destinations is enabled, e.g., both source and destination can talk to each other.  The interaction can come in two ways:  1) {\em In-band interaction:} sources and destinations can transmit and listen in the same channel simultaneously, enabling interaction. 2) {\em out-of-band interaction:} destinations talk back to the sources on an out-of-band channel, possible from white-space channels.

Although \cite{CJ2009} shows that for interference channel, relays, feedback, and full-duplex operation cannot improve the degrees of freedom beyond $\frac{K}{2}$,   we demonstrate that the interaction among sources and destinations enables flexibility in designing simple interference alignment scheme and in several cases achieves the optimal degrees of freedom. Both of these interaction methods are enabled by full-duplex radios, especially the in-band interaction requires high quality full-duplex systems with good self-interference suppression, which have attracted renewed attention in recent times \cite{Sachin11} \cite{Duarte12}.

Our main contribution is to propose a simple interference alignment scheme by exploiting the interactions among sources and destinations, and prove that the scheme can achieve the optimal degrees of freedom for several classes of interference channels, including $3$-user  IFC with out-of-band interaction,  $3$-user and $4$-user IFC with in-band interaction, and $4$-user MIMO IFC with in-band interaction.  One specific aspect of our model, namely, feedback using the reciprocal interference channel, has been considered in prior work \cite{changho}. In this work, we improve on this state-of-the-art in two ways: we prove new results for this specific model and also generalize this model to exploit more general modes of interaction, which admits the possibility of source-cooperation \cite{SourceCooperation,SourceCooperation2}, destination-cooperation \cite{DestCooperation,DestCooperation2} and in-band feedback in a single setting, in order to achieve interference alignment. The general modes of interaction permit simpler schemes.  In particular, for $K=3,4$, we show surprisingly that in-band interaction permits the first practical scheme that can achieve $\frac{K}{2}$ degrees of freedom, even after accounting for interaction cost. In addition to these results, we do extensive numeric simulations and show the proposed interactive interactive alignment scheme also works for some other classes of IFC empirically. We use tools from algebraic geometry to show why success of numeric simulations can suggest that the scheme should work well for almost all channel parameters in a rigorous way.

Along the way, we present a mathematical method for understanding the degrees of freedom in a general network with cooperation, feedback and relaying, where the nodes are constrained to using linear schemes of a fixed block length. While in general, the degrees of freedom achievable in a network will depend on the channel realization, in the case of all known channels, the degrees of freedom is the same for a measure $1$ of the channel realizations; and hence this value is called the degrees of freedom of the network.  We show that this remains true for general networks with linear schemes as well. Furthermore, once the block length is fixed we give a numerical way of estimating this number.

The paper is organized as follows. We describe the channel model in Section \ref{sec:model} and give some motivating examples in Section \ref{sec:specialK}. We present the interactive communication scheme for out-of-band interaction and write down the interference alignment conditions in Section \ref{sec:problemformulation}, and  our main technical results from algebraic geometry on the interference alignment feasibility in Section \ref{sec:math}. We also show how to use this method to infer properties of degrees of freedom of a general network with linear schemes in Section \ref{sec:math}. Section \ref{sec:scheme} and Section \ref{sec:fullduplex}
study interactive interference alignment for $K$-user interference channels with out-of-band interaction and in-band interaction, respectively. Simulation results on the finite SNR performance evaluation are presented in Section \ref{sec:simulation}. Section \ref{sec:discussion} discusses a general multi-phase interactive communication scheme for  $K$-user interference channel with large $K$, and lists several open problems. Section \ref{sec:conclusion} concludes this paper.

\section{System Model} \label{sec:model}

Consider a $K$-user interference channel with $2K$ radios, where radios $s_1,s_2,\dots,s_K$ are sources and radios $t_1,t_2, \dots, t_K$ are destinations. For each $1 \le i \le K$, source $s_i$ wants to send an independent message $W_i$ to destination $t_i$.

Let $H \in \C^{K \times K}$ denote the forward channel matrix from sources to destinations, and the input and output signals of the forward channel are related as
\begin{align}
y_i[n] = \sum_{j=1}^{K} H_{ij} x_j[n] + z_i [n],
\end{align}
where $n$ is the time index, $y_i[n]$ is the signal received by destination $t_i$ at time $n$, $x_j[n]$ is the signal sent out by source $s_j$ at time $n$, $z_i[n]$ is the channel noise, and $H_{ij}$ is the channel coefficient from source $s_j$ to destination $t_i$.

The above is a canonical channel model for $K$-user interference channel.

In this work, we consider two channel models where interaction among   sources and destinations can be enabled.

\subsection{Out-of-Band Interaction}
The first model we consider is a simple model for out-of-band interaction. While the sources talk to the destinations on its channel, the destinations are assumed to talk back to the source in a different channel, which could come from white-space channel.  Depending on how the out-of-band channel is obtained, one may or may not want to account for the cost of this channel. Here we will see that when $K\geq4$, this mode is useful even in the absence of 
Let $G \in \C^{K \times K}$ denote the feedback channel (or reverse channel) matrix from destinations to sources. Similarly, the input-output relation  of the feedback channel is
\begin{align}
f_i[n] = \sum_{j=1}^{K} G_{ij} v_j[n] + \tilde{z}_i [n],
\end{align}
where $n$ is the time index, $v_j[n]$ is the signal sent out by destination $t_j$ at time $n$, $f_i[n]$ is the signal received by source $s_i$ at time $n$,  $\tilde{z}_i[n]$ is  the channel noise, and $G_{ij}$ is  channel coefficient from destination $t_j$ to source $s_i$. 

This mode can also be enabled by in-band interaction using half-duplex radios. If the destinations use in-band half-duplex transmissions, due to the reciprocity of wireless channels, we have
\begin{align}
 G = H^T,
\end{align}
where $H^T$ denotes the transpose of $H$.

\begin{figure}[t]
\centering
\includegraphics[scale=0.5]{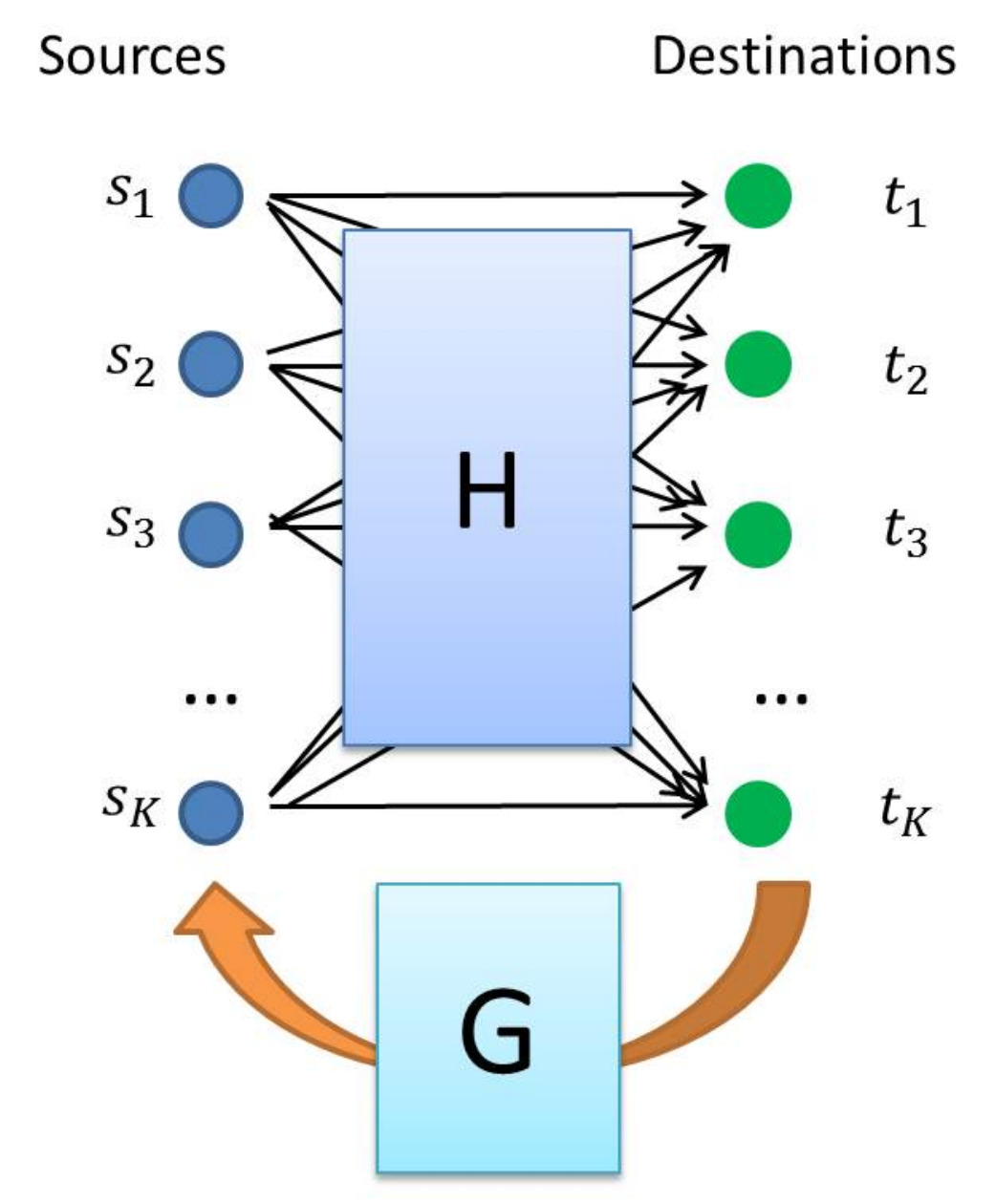}
\caption{System Model For Out-of-Band Interaction: $s_1,s_2,\dots,s_K$ are sources   and $t_1,t_2,\dots,t_K$  are destinations. $H$ is the forward channel matrix from sources to destinations, and $G$ is the feedback channel matrix from destinations to sources.}
\end{figure}

\subsection{In-Band Interaction}

Full-duplex radios can both send and receive signals using the same channel simultaneously, and this capability naturally enables the interaction among all radios in the network. We assume that all sources and destinations have full-duplex antennas, and all nodes can transmit and receive signals using the same channel (in-band) simultaneously. Let $H \in \C^{K \times K}$ be the channel matrix from sources to destinations,  $U \in \C^{K \times K}$ be the channel matrix among sources, and  $W \in \C^{K \times K}$ be the channel matrix among destinations. The input-output relation of this interference channel with in-band interaction is
\begin{align}
y_i[n] &= \sum_{j=1}^{K} H_{ij} x_j[n] + \sum_{j=1}^{K} W_{ij} v_j[n]  + z_i [n], \\
f_i[n] &= \sum_{j=1}^{K} H_{ji} v_j[n] + \sum_{j=1}^{K} U_{ij} x_j[n]  + \tilde{z}_i [n],
\end{align}
where $n$ is the time index, $x_j[n]$ and  $v_j[n]$ are the signals sent out by source $s_j$ and destination $t_j$ at time $n$, respectively,  $y_i[n]$ and $f_i[n]$ are the signal received by destination $t_i$ and source $s_i$ at time $n$,  $z_i [n]$ and $\tilde{z}_i[n]$ are  the channel noise,  $H_{ij}$ is  channel coefficient from   $s_j$ to   $t_i$ and also the channel coefficient from $t_i$ to $s_j$ due to channel reciprocity, $W_{ij}$ is the channel coefficient from $r_j$ to $r_i$, and $U_{ij}$ is the channel coefficient from $s_j$ to $s_i$.

\begin{figure}[t]
\centering
\includegraphics[scale=0.5]{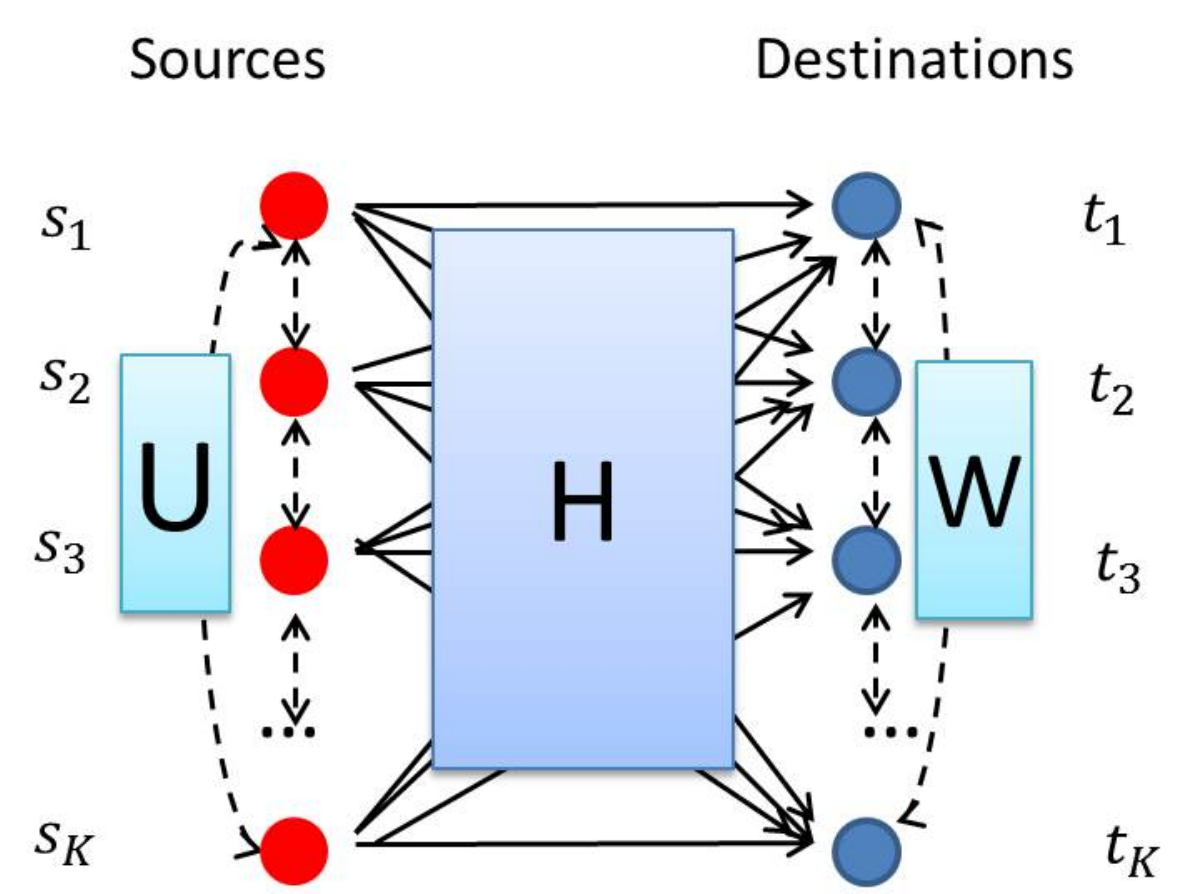}
\caption{System Model For In-Band Interaction: $s_1,s_2,\dots,s_K$ are sources and $t_1,t_2,\dots,t_K$  are destinations. $H$ is the forward channel matrix from sources to destinations,   $U$ is the   channel matrix among  sources, and $W$ is the channel matrix among destinations.}
\end{figure}

\section{Motivating Examples}\label{sec:specialK} 

The main theme of this paper is to show how interaction among destinations and sources can help do interference alignment/neutralization and thus make concurrent wireless communication feasible.

In this section, we start with a simple example of 3-user  interference channel  with out-of-band interaction, where communication is impossible without interaction from destinations, and describe a simple three-phase transmission scheme which makes concurrent communication feasible by exploiting the interaction from destinations to sources. Then, we generalize this three-phase transmission scheme to a class of $K$-user interference channels where the forward channel matrix $H$ can be written as the sum of a diagonal matrix and a rank 1 matrix.

\subsection{A symmetric 3-user interference channel}

Consider a 3-user interference channel with forward channel matrix
\begin{align}	
  H = \left(
  \begin{array}{ccc}
    0 & 1 & 1 \\
    1 & 0 & 1 \\
    1 & 1 & 0 \\
  \end{array}
\right),
\end{align}
where all interference link gains are 1, and all direct link gains are 0.

Since the direct links have gains 0, it is trivial to see that without interaction from destinations, no reliable communication is possible. However, if we allow destinations to talk back to sources using the reciprocal channel, we can  achieve $\frac{3}{2}$  degrees of freedom as seen below. For simplicity, let $G=H^T$ be the reverse channel; one way of achieving this is by using half-and-half interaction.  

Consider the following three-phase interactive transmission scheme:
\begin{itemize}
  \item Phase 1: All sources send their independent symbols $x_1, x_2, x_3$ simultaneously. Then destinations receive
  \begin{align}
     y_1 &= x_2 + x_3 + \text{noise}, \\
     y_2 &= x_1 + x_3 + \text{noise}, \\
     y_3 &= x_1 + x_2 + \text{noise}.
   \end{align}

  \item Phase 2: After receiving signal $y_1, y_2, y_3$ from sources in Phase 1, all destinations simultaneously send out $-y_i$ to sources using the reciprocal channel, and sources  get
  \begin{align}
     f_1 &= -y_2 - y_3 + \text{noise}  \nonumber \\
         &= -2x_1 - x_2 - x_3 + \text{noise}, \\
     f_2 &= -y_1 - y_3 + \text{noise} \nonumber \\
     &= -2x_2 - x_1 - x_3 + \text{noise}, \\
     f_3 &= -y_1 - y_2 + \text{noise}\nonumber \\
      &= -2x_3 - x_1 - x_2 + \text{noise}.
   \end{align}

  \item Phase 3: Now each source $s_i$ has two sets of signals $x_i$ and $f_i$, and each $s_i$  can simultaneously send out   $(-3 x_i - f_i)$ to destinations using the forward channel. destinations will get
   \begin{align}
     y'_1 &= (-3 x_2 - f_2) + (-3 x_3 - f_3) + \text{noise} \nonumber \\
     &= 2 x_1 + \text{noise}, \\
     y'_2 &= (-3 x_1 - f_1) + (-3 x_3 - f_3) + \text{noise} \nonumber \\
     &= 2 x_2 + \text{noise}, \\
     y'_3 &= (-3 x_1 - f_1) + (-3 x_2 - f_2) + \text{noise} \nonumber \\
     &= 2 x_3 + \text{noise}.
   \end{align}
\end{itemize}

Therefore, after two forward transmissions and one reverse transmission, each destination $t_i$ will get the desired symbol $x_i$ from the corresponding source $s_i$ without any interference. If we do not count the reverse transmission, the total degrees of freedom achieved are $\frac{3}{2}$. Even if we account for reverse transmission, we get a total degrees of freedom of $1$, which cannot be achieved in the absence of

\subsection{A symmetric $K$-user interference channel with special channel matrix}

We can extend the above result to $K$-user interference channel where the channel matrix is an all-one matrix except all diagonal entries being zero for any $K \ge 3$.

Consider a $K$-user interference channel where the channel matrix $H$ satisfies
\begin{align}
  H_{ij} &= 1, \forall i \neq j \\
  H_{ii} &= 0, \forall 1 \le i \le K.
\end{align}

We can show that the following three-phase interactive transmission scheme with two forward transmissions and one reverse transmission can make  each destination $t_i$ get the desired symbol from source $s_i$ for all $1 \le i \le K$.

\begin{itemize}
  \item Phase 1: All sources send their independent symbols $x_1, x_2, \dots, x_K$ simultaneously. Then each destination  $t_i$ get
  \begin{align}
     y_i &= \sum_{1 \le j \le K, j \neq i} x_j + \text{noise}.
   \end{align}

  \item Phase 2: After receiving signal $y_1, y_2, \dots, y_K$ from sources in Phase 1, all destinations simultaneously send out $-y_i$ to sources using the reciprocal channel. Then each source $s_i$  gets
  \begin{align}
     f_i &= \sum_{1 \le j \le K, j \neq i} y_j + \text{noise} \nonumber \\
     &= -(K-1)x_i - \sum_{1 \le j \le K, j \neq i} (K-2)x_j   + \text{noise}.
   \end{align}

  \item Phase 3: Now each source $s_i$ has two sets of signals $x_i$ and $f_i$, and each $s_i$  can simultaneously send out
       \begin{align}
       ( -(K^2-3K+3) x_i - f_i)
       \end{align}
       to destinations using the forward channel. Then each destination $t_i$ will get
   \begin{align}
     y'_i &= \sum_{1 \le j \le K, j \neq i} ( -(K^2-3K+3) x_i - f_i) + \text{noise} \nonumber \\
     &= (K-2)(K-1) x_i + \text{noise}.
   \end{align}
\end{itemize}

For $K \ge 3 $, $(K-2)(K-1) \neq 0 $. Therefore, each destination $t_i$ will get the desired symbol from the source $s_i$ without any interference for all $1 \le i \le K$.

In fact, we can show that if $H \in \C^{K \times K}$ can be written as the sum of diagonal matrix and a rank 1 matrix, then in general two forward transmissions and one reverse transmission can make each destination $t_i$ get the desired symbol from source $s_i$ for all $1 \le i \le K$.

\begin{theorem}\label{thm:special}
Consider a special $K$-user interference channel with reciprocal feedback channel ($G = H^T$), where the forward channel matrix $H$ can be written as the sum of a diagonal matrix and a rank 1 matrix, i.e.,
\begin{align}
	H = D + uv^T,
\end{align}
where $D \in \C^{K\times K}$ is a diagonal matrix and $u,v \in \C^{K \times 1}$ are  column vectors.

If $D$ is invertible, each component of $u$ and $v$ is nonzero, and
\begin{align}
	 \alpha \triangleq	3+ 3v^TD^{-1}u + (v^TD^{-1}u )^2 \neq 1,
\end{align}
	then there exists a three-phase interactive transmission scheme such that  two forward transmissions and one reverse transmission can make each destination $t_i$ get the desired symbol from source $s_i$ without any interference for all $1 \le i \le K$.
\end{theorem}

\begin{IEEEproof}
See Appendix \ref{app:special}.
\end{IEEEproof}


\section{Communication Scheme and Interference alignment conditions }\label{sec:problemformulation} 

The three-phase interactive transmission scheme described in Section \ref{sec:specialK} is a natural scheme to exploit the interaction from destinations  to sources to make concurrent wireless communication feasible.  We can generalize the transmission scheme there to any interference channel with out-of-band interaction. We will do so  in this section, by first describing the scheme with certain design parameters and then writing down the constraints required on the design parameters in order for interference alignment to be achieved.   

\begin{itemize}
  \item Phase 1 (forward transmission): All sources send their independent symbols  simultaneously. And destinations get $y = H x + n$, where $n$ is the  additive noise of the channel.
  \item Phase 2 (interaction from destinations): After receiving signals from sources in phase1, all destinations scale $y$ and send back to sources using the reverse channel. Sources  get $f = GD_1 y +  \tilde{n}$, where $\tilde{n}$ is the  additive noise of the channel. Since each source  and each destination only knows what signals they sent out and received, the coding matrix  $D_1$ has to be diagonal.
  \item Phase 3 (forward transmission): Now each source has two sets of signals $x$ and $f$, and sources can send out a linear combination of $x$ and $f$ to destinations via the forward channel. Destinations get
  \begin{align}
    y' &= H(D_2 x + D_3 f) + \hat{n} \\
      & =  (HD_2 + HD_3 G D_1H) x +  HD_3GD_1 n + HD_3 \tilde{n} + \hat{n}.
  \end{align}
  Again, the matrices $D_2,D_3$ are constrained to be diagonal.
\end{itemize}

\begin{figure}[t]
\centering
\includegraphics[scale=0.5]{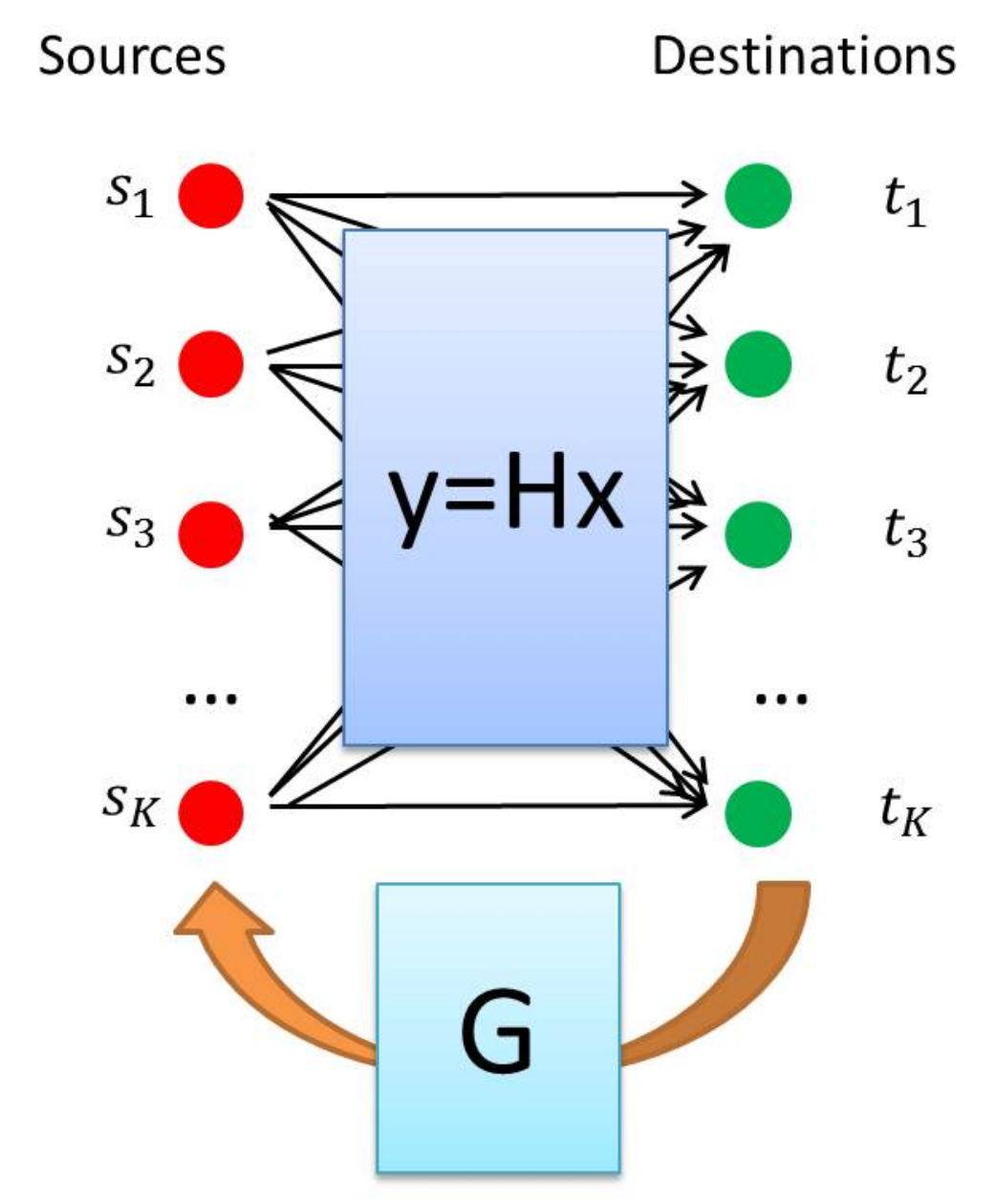}
\caption{Phase 1: All sources send their independent symbols  simultaneously, and destinations get $y = H x + \text{noise}$.}
\end{figure}

\begin{figure}[t]
\centering
\includegraphics[scale=0.5]{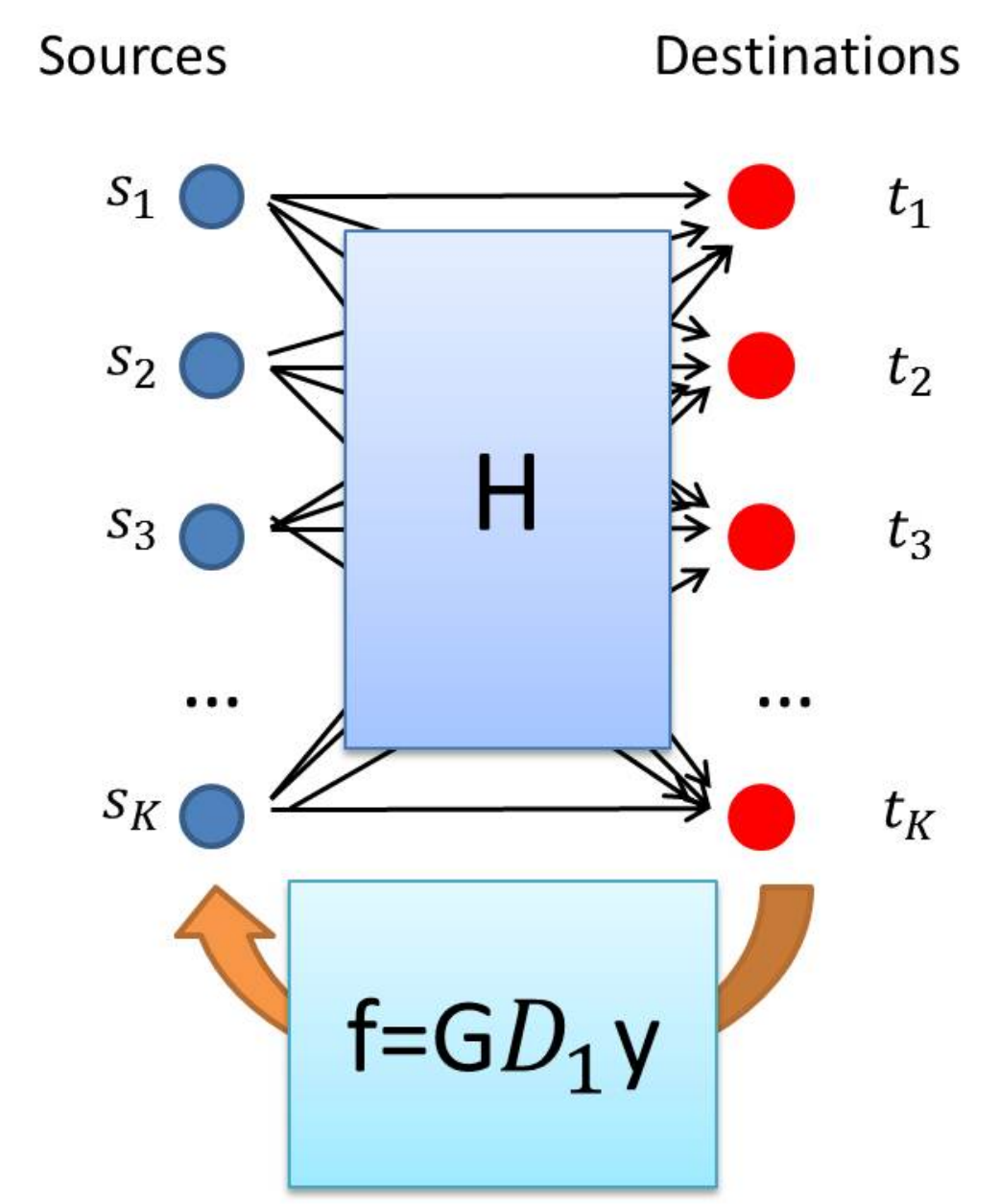}
\caption{Phase 2: After receiving signals from sources in phase1, all destinations scale $y$ and send back to sources using the feedback channel. Sources  get $f = GD_1 y +  \text{noise}$.}
\end{figure}

\begin{figure}[t]
\centering
\includegraphics[scale=0.5]{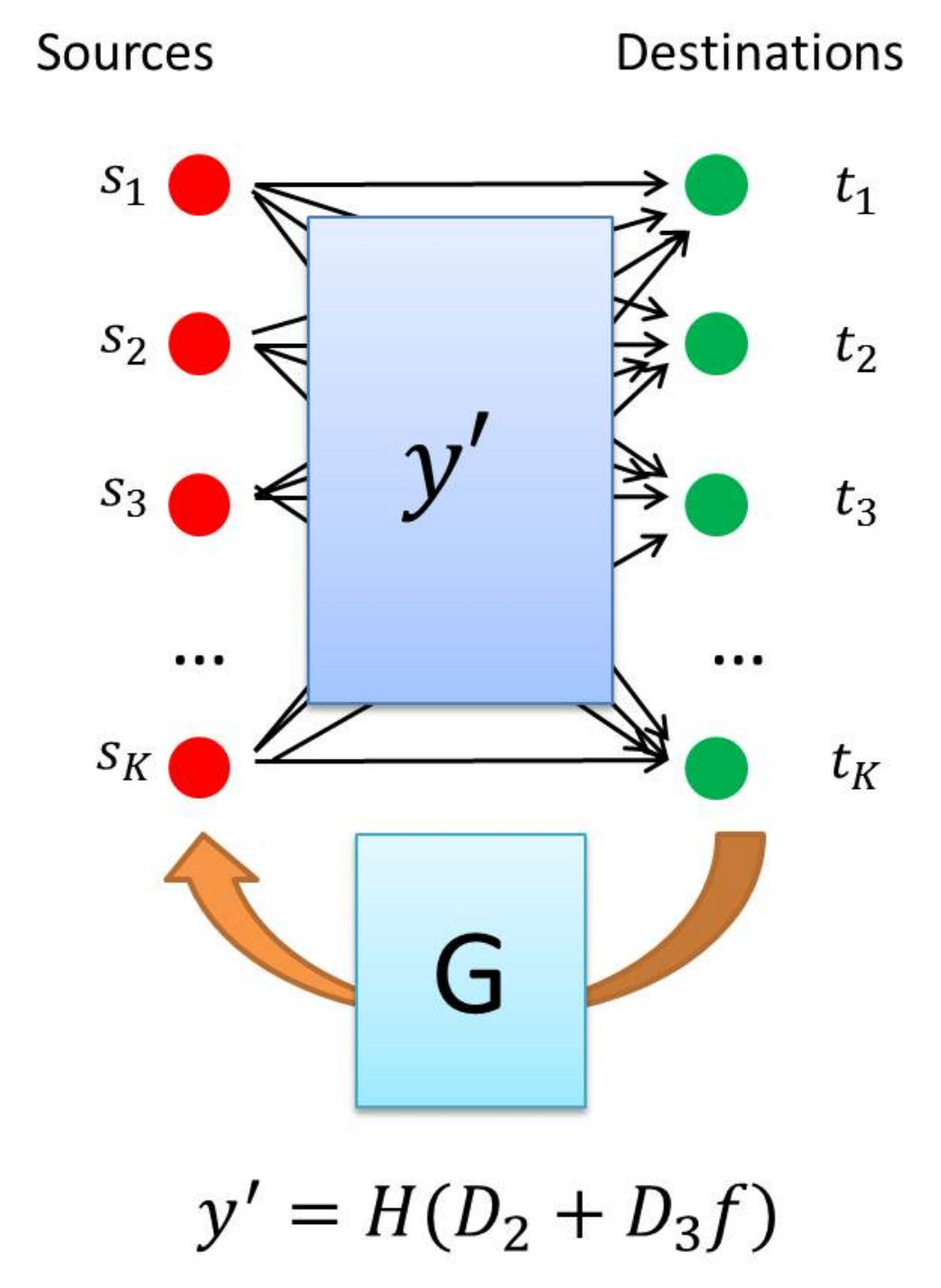}
\caption{Phase 3: Now each source has two sets of signals $x$ and $f$, and sources can send out a linear combination of $x$ and $f$ to destinations via the forward channel. Destinations get $y' = H(D_2 x + D_3 f) +  \text{noise}$.}
\end{figure}


In the above scheme, since each source  and each destination only know what signals they sent out and received, the coding matrices  $D_1, D_2$ and $D_3$ have to be diagonal, i.e., each node can only do coding over what signals it sent out and received. Note that since we are interested only in degrees of freedom calculations in this paper, we will not keep track of the particular structure of noise, as long as it has finite variance and is independent of everything else in the equation.

To make destinations be able to decode desired signals, one possibility is do interference neutralization, i.e., the interference term is zero for each of the receivers. This can be guaranteed if we choose diagonal matrices $D_1, D_2$ and $D_3$ such that
$(HD_2 + HD_3 G D_1H )$ is a diagonal matrix and every diagonal entry is nonzero, which makes each destination get the desired signal from $y'$ without any interference. In fact we have  used coding matrices with this property   for the special class of interference channels presented in Section \ref{sec:specialK}. So a natural question to ask is whether there always exists such $D_1,D_2$ and $D_3$.

Requiring that all off-diagonal entries be zero gives rise to $K(K-1)$ polynomial equations, and each diagonal entry being nonzero leads to $K$ inequalities, whereas in total there are only $3K$ variables, corresponding to the $K$ variables in each of the diagonal coding matrices $D_1, D_2$ and $D_3$. One may therefore be led to conjecture that if the number of  variables is more than the number of equations, a solution should exist that simultaneously satisfies all equations and inqualities. Unfortunately, this conjecture is not true: there are cases where even though the number of variables is more than equations, due to the coupled constraints on the equations and inequalities, all off-diagonal entries of $(HD_2 + HD_3 G D_1H )$ being zero implies that at least one diagonal entry is zero. A concrete example is the 3-user interference channel with reciprocal reverse channel, which will be discussed in Section \ref{sec:scheme}.

One way to overcome this problem is to do interference alignment. In this case, we want to align the interference seen by the receiver during the first and the third phase, i.e., the interference terms in $y$ (which the destinations received during the first phase) and $y'$ (which destinations receive during the third phase) instead of relying purely on $y'$ alone, which was the case in interference neutralization.  Now, the destinations can cancel out all interference by taking a linear combination of $y$ and $y'$ if the interference terms in $y'$ are aligned with the interference terms in $y$.

More precisely, let $ B \triangleq HD_2 + HD_3 G D_1H $. Then for destination $t_i$, in Phase 1 it receives
\begin{align}
  y_i = H_{ii} x_i + \sum_{j \neq i} H_{ij} x_j + \text{noise},
\end{align}
and in Phase 3 it receives
\begin{align}
  y'_i = B_{ii} x_i + \sum_{j \neq i} B_{ij} x_j + \text{noise}.
\end{align}

If for all $1 \le i \le K$, we have
\begin{align}
  \frac{B_{ij}}{H_{ij}} = \frac{B_{ij'}}{H_{ij'}} , \quad \forall j,j'\neq i, \label{eqn:key1}
\end{align}
or equivalently
\begin{align}
  \frac{B_{ij}}{H_{ij}} = \lambda_i, \quad \forall j \neq i,
\end{align}
for some constant $\lambda_i$,
then interference terms of $y$ and $y'$ for each destination are aligned with each other. So destination $t_i$ can compute $y_i - \lambda_i y'_i$ and get
\begin{align}
   (H_{ii} - \lambda_i B_{ii})x_i + \text{noise}.
\end{align}
Therefore, if $(H_{ii} - \lambda_i  B_{ii}) \neq 0$, or equivalently,
\begin{align}
  \frac{B_{ii}}{H_{ii}} \neq \frac{B_{ij}}{H_{ij}}, \quad \forall j \neq i, \label{eqn:key2}
\end{align}
then destination $t_i$ can get the desired signal sent by source $s_i$ without any interference.

Without introducing the auxiliary variables $\lambda_i$, aligning $(K-1)$ interference terms for each destination corresponds to $(K-2)$ equations, and thus in total we have $K(K-2)$ interference alignment equations. Preserving the desired signals after canceling all interference terms leads to $K$ inequalities. Therefore, to make the three-phase communication scheme work and thus achieve the optimal $\frac{K}{2}$ degrees of freedom, we need to solve $K(K-2)$ equations with $3K$ variables and check that whether the solution satisfies  the $K$ inequalities.

In some cases, we can reduce the polynomial equations corresponding to interference alignment equality constraints to linear equations, the solution of which has a closed-form expression and thus it can be easily verified whether the inequality constraints are satisfied. However, in general the system of polynomial equations is nonlinear, and it may not have closed-form solution, making it hard to check whether the inequality constraint can be satisfied. Our main approach is to convert the system of polynomial equations and polynomial inequalities to a system of polynomial equations, and then use tools from algebraic geometry to check the existence of solution for the system of polynomial equations.

First, we show how to reduce the problem to checking the existence of solutions to a system of polynomial equations. Then we will present our main technique for solving the polynomial system using tools from algebraic geometry in Section \ref{sec:math}.

Suppose there are $N$ polynomial equation constraints $f_i = 0$, $1 \le i \le N $, and $M$ polynomial inequality constraints $g_j \neq 0$, $1 \le j \le M$, over $S$ variables $\{d_1,d_2, \dots, d_S\}$. By introducing an auxiliary variable $t$, we define a polynomial function $\hat{g}$ as
\begin{align}
   \hat{g} \triangleq t\prod_{j=1}^M g_j - 1.
 \end{align}

 \begin{lemma}\label{lem:equivalence}
   There exists $(d_1,d_2, \dots, d_S)$ satisfying equality and inequality constraints
   \begin{align}
     f_i = 0 , \quad \forall 1 \le i \le N \label{eqn:equalities}\\
     g_j \neq 0, \quad \forall 1 \le j \le M, \label{eqn:inequalities}
   \end{align}
   if and only if there exists $(d_1,d_2, \dots, d_S, t)$ satisfying
    \begin{align}
     f_i = 0 , \quad \forall 1 \le i \le N \label{eqn:newequalities}\\
     \hat{g} = 0. \label{eqn:newone}
   \end{align}
 \end{lemma}

\begin{IEEEproof}
  If $(d_1,d_2, \dots, d_S)$ is a solution to \eqref{eqn:equalities} and \eqref{eqn:inequalities}, then
  $(d_1,d_2, \dots, d_S, t)$ is a solution to \eqref{eqn:newequalities} and \eqref{eqn:newone}, where
  \begin{align}
    t \triangleq \frac{1}{\prod_{j=1}^M g_j}.
  \end{align}

  On the other hand, if $(d_1,d_2, \dots, d_S, t)$ is a solution to \eqref{eqn:newequalities} and \eqref{eqn:newone}, then $(d_1,d_2, \dots, d_S)$ satisfies both \eqref{eqn:equalities} and  \eqref{eqn:inequalities}.
\end{IEEEproof}

Therefore, due to Lemma \ref{lem:equivalence}, the problem of checking the existence of solutions to a system of polynomial equations and inequalities can be reduced to the problem of checking existence of solutions of a system of polynomial equations, which is well studied for algebraically closed field in algebraic geometry \cite{AGbook}.

\section{General Solution Methodology}\label{sec:math} 

 In this section we present our main technical results on checking the existence of solutions to the interference alignment equations for generic channel parameters $H$ and $G$ using tools in algebraic geometry. For more details on algebraic geometry, we refer the reader to Appendix \ref{app:1} and the excellent textbook \cite{AGbook}.

We follow the standard notation in \cite{AGbook}.  Let $k$ denote a field, and let $k[\xi_1,\xi_2,\dots,\xi_n]$ denote the set of all polynomials in the variables $\xi_1,\xi_2,\dots,\xi_n$ with coefficients in $k$.

\begin{definition}[Definition 1 on page 5 of \cite{AGbook}]
  Let $k$ be a field, and let $f_1,\dots,f_s$ be polynomials in $k[\xi_1,\xi_2,\dots,\xi_n]$. Then we call
  \begin{align}
    V(f_1,\dots,f_s) \triangleq \{ (a_1,\dots,a_n)\in k^n \ |\  & f_i(a_1,\dots,a_n) = 0, \nonumber \\
     & \quad \forall 1 \le i \le s\}
  \end{align}
  the \textbf{affine variety} defined by $f_1,\dots,f_s$.
\end{definition}

As discussed in Section \ref{sec:problemformulation}, our problem can be reduced to checking the existence of solutions to a system of polynomial equations. In the language of algebraic geometry, the problem is to check whether the affine variety defined by some polynomials is an empty set or not, which is well studied  for algebraically closed field in algebraic geometry. In wireless communication, the channel coefficients are represented as complex numbers $\C$, which is an algebraically closed field.

The standard approach to checking whether an affine variety is an empty set  is to use Buchberger's algorithm to compute the Gr\"{o}bner basis of the given polynomials, and from the Gr\"{o}bner basis we can easily conclude whether the corresponding affine variety is empty or not \cite{Buchberger}.

One important implication of these results in algebraic geometry is that  if the coefficients in the polynomial equations are rational functions of variables $\{h_i\}$, then  except a set of $\{h_i\}$ which satisfies a nontrivial polynomial equation on $\{h_i\}$ and thus has a measure 0,  either for all numeric values of $\{h_i\}$ the polynomial equations have a solution, or for all numeric values of $\{h_i\}$ the polynomial equations do not have a solution.

More precisely,

\begin{theorem}\label{thm:equivalsym_num}
  Let $f_1,\dots,f_s$ be polynomials in $k[\xi_1,\xi_2,\dots,\xi_n]$, where $k = \C$ and all coefficients of the polynomials are rational functions of variables $h_1,h_2,\dots,h_m$. Then there exists a nontrivial polynomial equation on $h_1,h_2,\dots,h_m$, denoted by $e(h_1,h_2,\dots, h_m)$, such that except the set of $\{(h_1,h_2,\dots,h_m)\  |\  e(h_1,h_2,\dots,h_m) = 0 \}$, either for all $ (h_1, h_2, \dots, h_m) \in \C^{1 \times m} $, $V(f_1,\dots,f_s) \neq \emptyset$, or for all $(h_1, h_2, \dots, h_m) \in \C^{1 \times m}$, $V(f_1,\dots,f_s) = \emptyset$.
\end{theorem}
\begin{IEEEproof}
See Appendix \ref{app:proof}.
\end{IEEEproof}

For the polynomial equations describing the interference alignment problem, the coefficients of the polynomials are rational functions of channel parameters $H$ and  $G$ in symbolic form. Therefore, in the context of the interference alignment feasibility problem, this main result can be restated as follows. {\em Either one} of the following two statements hold:
\begin{itemize} \item For almost all\footnote{We emphasize that in our results ``almost all'' means for all numeric values except a set  of parameters which satisfy a nontrivial polynomial equation. Therefore, in contrast to results in \cite{Wu2011} and \cite{RealInter2009}, our results are not sensitive to whether the channel parameters are rational or irrational.} channel realizations of $H$ and $G$, there exists solution to the system of interference alignment equations.
\item Or, for  almost all channel realizations of $H$ and $G$, there does not exist solution to the system of interference alignment equations. \end{itemize}

Although in theory we can compute in finite number of steps the symbolic Gr\"{o}bner basis for these polynomials with symbolic coefficients to check whether for almost all $H$ and $G$ there exists solutions, it turns out to be computationally infeasible to run the Buchberger's algorithm for the symbolic polynomial equations for most of our interference alignment problems, due to the fact that the orders of intermediate symbolic coefficients can increase exponentially.
However, it is much easier to compute a Gr\"{o}bner basis numerically. Due to Theorem \ref{thm:equivalsym_num},
\begin{corollary}\label{cor:key}
  If we draw the channel parameters according to a continuous probability distribution, then with probability one the numeric polynomial equations have a  solution if and only if for almost all channel realizations the polynomials equations have solution.
\end{corollary}

Hence, while we may not be able to prove that  certain polynomial equations have solution for almost all channel parameters because of computational difficulty, numeric simulations can let us make claims with high credibility.

\subsection{Application to General Networks}
In a general network with potentially feedback, cooperation and relaying, an important question to understand is when linear schemes can acheive a certain degrees of freedom.  While this is a hard question in general, once we restrict to a fixed block length, we can start answering this question. Let us fix a block length of communication and let each user transmit a linear combination of the symbols that he received. At the end of this communication, the receivers apply a linear combination of all received inputs to construct the decoded vector. We want this decoded vector to equal the transmitted vector. We leave all the multiplication matrices to be design variables and ask when does this system of equations have a solution. The answer to this question, of course, depends on the realization of the channel.  However, if we assume that that channel coefficients are drawn from a measure with a probability density, then we can invoke Theorem \ref{thm:equivalsym_num} and Corrolory~\ref{cor:key} to show that there are only two possibilities.
\begin{enumerate} 
\item With probability $1$ over the channel measure, the particular degrees of freedom is achievable.
\item With probability $1$ over the channel measure, the particular degrees of freedom is not achievable. 
\end{enumerate}

In order to test which of the two hypotheses is true, we can run simulations, where the channel is generated according to the measure. Once the channel is fixed, we can then run numerical Gr\"{o}bner basis algorithm to determine whether there is a solution or not. If for almost all the simulations, the solution exists, then we can declare that the former case is true; if not, the latter case is true. While this step of going from the simulation to the conclusion has technical difficulties due to the fact that computer has numerical precision, we can make such claims with high confidence.

\section{Out-of-Band Interactive Interference Alignment}\label{sec:scheme}
In this section, we study interactive interference alignment for $K$-user  interference channel with out-of-band interaction. We show results only for small values of $K$, in particular only for $K \leq 4$.

\subsection{$K = 3$}
We described in Section \ref{sec:problemformulation}, a three-phase interactive communication scheme for out-of-band interference alignment and the required conditions on the design matrices. Recall that, for such a scheme work, we would like to find diagonal coding matrices $D_1$, $D_2$ and $D_3$ to make $(HD_2 + HD_3 G D_1H)$ be a diagonal matrix with each diagonal entry being nonzero, so that interference neutralization can be achieved in the third phase. For $3$-user interference channel, the number of interference neutralization equations is $6$, and the number of variables is $3K = 9$. Even so, one may hope that the interference neutralization equations are solvable, due to the coupled constraints on the equations and inequalities, it can be quickly seen that all off-diagonal entries of $(HD_2 + HD_3 G D_1H )$ being zero implies that at least one diagonal entry is zero. A simple counter example is the case when the reverse channel is reciprocal to the forward channel, i.e., $G = H^T$. We did extensive numeric simulations, and in each numeric instance we find that there is no solution to solve both equations and inequalities by using the methodology (computing Gr\"{o}bner basis)  introduced in section \ref{sec:math}. Therefore, due to Corollary \ref{cor:key}, we believe the following claim holds.

\begin{claim}
  For 3-user interference channel with reciprocal feedback channel ($G = H^T$), for generic matrix\footnote{Generic matrix means all numeric matrices except a set of matrices the entries of which satisfy a nontrivial polynomial equation.}  $H$, there does not exist $D_1,D_2$ and $D_3$ such that $(HD_2 + HD_3 G D_1H )$ is a diagonal matrix and every diagonal entry is nonzero.
\end{claim}

As discussed in Section \ref{sec:problemformulation}, it is not necessary to do exact interference neutralization. Instead, interference alignment is also sufficient. When $K=3$,  the number of variables is $3K = 9$, which is much more than the number of interference alignment equations $K(K-2) = 3$, so it is not surprising that we can find  $D_1, D_2$ and $D_3$ to solve the interference alignment equations and inequalities. \cite{changho} gives an analytical solution to do interference alignment for $3$-user interference channel with feedback. Here, we present a different proof technique, which also will be applied in interference alignment with in-band interaction in Section \ref{sec:fullduplex}.

First we state a lemma which will be used in the proof of Theorem \ref{thm:3user}.

\begin{lemma}\label{lem:dimension}
  Let $S_0 \subset \C^n$ be a linear subspace with dimension $d_0$. For any $m$ linear subspaces $S_1,S_2,\dots,S_m \subset S_0$ with dimensions $d_1,\dots,d_m$ respectively, if
  \begin{align}
    d_i < d_0, \quad \forall 1 \le i \le m,
  \end{align}
  then $S_0 \backslash (\cup_{i=1}^m S_i)$ is a nonempty set.
\end{lemma}

\begin{IEEEproof}
  Let $e_1,\dots,e_{d_0} \in \C^n $ be a set of bases of $S_0$. Let $A \in \C^{d_0 \times d_0 m}$ such that every $d_0$ by $d_0$ submatrix of $A$ has full rank. It is easy to show the existence of $A$ with such property. Indeed, if every entry of $A$ is independently generated according to Gaussian distribution (or other continuous probability distributions), then with probability 1 every $d_0$ by $d_0$ submatrix of $A$ has full rank.

  Define $d_0 m$ vectors $b_1,b_2,\dots,b_{d_0 m}$ by setting
  \begin{align}
    b_i \triangleq \sum_{j=1}^{d_0} A_{ji} e_j,
  \end{align}
  for all $1 \le i \le d_0 m$.
  Since every $d_0$ by $d_0$ square submatrix of $A$ has full rank, every $d_0$ vectors of $b_1,b_2,\dots,b_{d_0 m}$ are linearly independent.

  Suppose $S_0 \backslash (\cup_{i=1}^m S_i)$ is the empty set, i.e., $S_0 = \cup_{i=1}^m S_i$. Then $b_1,b_2,\dots,b_{d_0 m} \in \cup_{i=1}^m S_i$. By pigeonhole principle, there exists $i^* \le d_0 m$ such that $S_{i^*}$ contains at least $d_0$ vectors of $b_1,b_2,\dots,b_{d_0 m}$. Since every  $d_0$ vectors of $b_1,b_2,\dots,b_{d_0 m}$ are linearly independent, the dimension of $S_{i^*}$ is no less than $d_0$, contradicting with the fact that $d_{i^*} < d_0$.

  Therefore, $S_0 \backslash (\cup_{i=1}^m S_i)$ is a nonempty set.

\end{IEEEproof}

\begin{theorem}\label{thm:3user}
  For 3-user interference channel with reciprocal feedback channel ($G = H^T$), and for generic channel matrix $H$, there exists solutions to the interference alignment equations and inequalities.
\end{theorem}

\begin{IEEEproof}
The $K(K-2) = 3$ equations are
\begin{align}
  \frac{B_{12}}{H_{12}} &= \frac{B_{13}}{H_{13}}, \label{eqn:linear1} \\
  \frac{B_{21}}{H_{21}} &= \frac{B_{23}}{H_{23}}, \label{eqn:linear2} \\
  \frac{B_{31}}{H_{31}} &= \frac{B_{32}}{H_{32}}, \label{eqn:linear3}
\end{align}
and the $K = 3$ inequalities are
\begin{align}
  \frac{B_{11}}{H_{11}} &\neq \frac{B_{12}}{H_{12}}, \label{eqn:nonlinear1}\\
  \frac{B_{22}}{H_{22}} &\neq \frac{B_{23}}{H_{23}}, \label{eqn:nonlinear2}\\
  \frac{B_{33}}{H_{33}} &\neq \frac{B_{32}}{H_{32}},\label{eqn:nonlinear3}
\end{align}

  Since the number of variables is much more than the number of equations, we do not need to use all the variables. In fact, we can set $D_1$ to be a simple diagonal matrix
\begin{align}
  D_1 = \left(
  \begin{array}{ccc}
    1 & 0 & 0 \\
    0 & 0 & 0 \\
    0 & 0 & 0 \\
  \end{array}
\right).
\end{align}

Write $D_2$ and $D_3$ as
\begin{align}
  D_2 = \left(
  \begin{array}{ccc}
    D_{21} & 0 & 0 \\
    0 & D_{22} & 0 \\
    0 & 0 & D_{23} \\
  \end{array}
\right), \\
 D_3 = \left(
  \begin{array}{ccc}
    D_{31} & 0 & 0 \\
    0 & D_{32} & 0 \\
    0 & 0 & D_{33} \\
  \end{array}
\right).
\end{align}

Then equations \eqref{eqn:linear1}, \eqref{eqn:linear2} and \eqref{eqn:linear3} are homogeneous linear equations in terms of variables $D_{21}, D_{22}, D_{23}, D_{31}, D_{32}$ and $D_{33}$. Let $S$ denote the set of solutions to linear equations \eqref{eqn:linear1}, \eqref{eqn:linear2} and \eqref{eqn:linear3}. It is easy to check that the corresponding coefficient matrix has a rank of $3$. Therefore, $S$ is a linear subspace of dimension $6 - 3 = 3$.

Consider the following equations corresponding to the inequality constraints
\begin{align}
  \frac{B_{11}}{H_{11}} &= \frac{B_{12}}{H_{12}}, \label{eqn:linear4}\\
  \frac{B_{22}}{H_{22}} &= \frac{B_{23}}{H_{23}}, \label{eqn:linear5}\\
  \frac{B_{33}}{H_{33}} &=\frac{B_{32}}{H_{32}},\label{eqn:linear6}.
\end{align}

Let $S_1$ denote the set of solutions to \eqref{eqn:linear1}, \eqref{eqn:linear2}, \eqref{eqn:linear3} and \eqref{eqn:linear4}, let $S_2$ denote the set of solutions to \eqref{eqn:linear1}, \eqref{eqn:linear2}, \eqref{eqn:linear3} and \eqref{eqn:linear5},  and let $S_3$ denote the set of solutions to \eqref{eqn:linear1}, \eqref{eqn:linear2}, \eqref{eqn:linear3} and \eqref{eqn:linear6}. We can check that all the coefficient matrices of the linear equations
 have a rank of $4$, so all of $S_1, S_2$ and $S_3$ are linear subspaces of dimension $6 - 4 =2$.

 Now consider the set $S^* \triangleq S \backslash (S_1 \cup S_2 \cup S_3)$. Any element in $S^*$  satisfies both the equations \eqref{eqn:linear1}, \eqref{eqn:linear2}, \eqref{eqn:linear3}, and the inequality constraints \eqref{eqn:nonlinear1}, \eqref{eqn:nonlinear2}, \eqref{eqn:nonlinear3}.

 Since $S$ is linear subspace of dimension $3$, which is larger than the dimensions of $S_1, S_2$ and $S_3$, $S^*$ is a nonempty set by Lemma \ref{lem:dimension}. This completes the proof.

\end{IEEEproof}


The same result continues to hold when the feedback channel matrix is also a generic matrix  independent of $H$.

It is worth pointing out that the 3-user interference channel under an alternate local feedback
model,  where each receiver sends feedback to its corresponding
transmitter, was studied in \cite{NC12} recently. In contrast to our model, where obtaining feedback using natural reciprocal channel takes up a single time slot, local feedback can take $3$ time slots to implement.

\subsection{$K = 4$}

For 4-user interference channel, we have $K(K-2) = 8$ polynomial equation constraints and $K = 4$ inequality constraints with $3K = 12$ variables. We cannot use the same technique as in the case of three-user interference channel by setting $D_1$ or $D_3$ to be a deterministic constant matrix and thus reducing the polynomial equations to linear equations. The reason is that if we do not use $D_1$ or $D_3$ as variables, the number of equations will be the same as the number of variables, and this will prevent us from finding a solution which also satisfies the inequality constraints.  Instead, we use the tools introduced in Section \ref{sec:math} to study the feasibility of interactive interference alignment. More precisely, we first convert the system of polynomial equations and polynomial inequalities to a system of polynomial equations, and then we check whether it has solutions or note by computing the  Gr\"{o}bner basis of corresponding polynomials.

Due to computational difficulty, it is hard to compute the Gr\"{o}bner basis for the symbolic polynomials. We do extensive numeric verifications: each time we generate random numeric channel matrices, and then apply Buchberger's algorithm to compute a  Gr\"{o}bner basis. In all numeric instances, the Gr\"{o}bner basis does not contain a nonzero scalar, which implies that solution to \eqref{eqn:key1} and \eqref{eqn:key2} exists. Therefore, due to Corollary \ref{cor:key}, although we cannot compute a symbolic Gr\"{o}bner basis explicitly, from numeric verifications we believe  the following claims hold.

\begin{claim}
  For 4-user interference channel with reciprocal feedback channel ($G = H^T$), for generic matrix $H$, there exists $D_1,D_2$ and $D_3$ such that each row of $(HD_2 + HD_3 G D_1H )$ is proportional to the corresponding row of $H$ except the diagonal entries.
\end{claim}

\begin{claim}
  For 4-user interference channel with out-of-band feedback channel, for generic matrices $H$ and $G$, there exists $D_1,D_2$ and $D_3$ such that each row of $(HD_2 + HD_3 G D_1H )$ is proportional to the corresponding row of $H$ except the diagonal entries.
\end{claim}

\cite{changho} formulates the interference alignment problem as a rank constrained nonconvex optimization problem, and reports that all numeric simulations confirm the existence of solutions to the interference alignment equations and inequalities.

In some cases when $H$ has special structures, we are able to compute the symbolic  Gr\"{o}bner basis of the interference alignment polynomials, and thus we can conclude whether for almost all $H$ with such structure there exists interference alignment solutions. One positive example is that if $H$ is a symmetric matrix and all diagonal entries are zero, then for almost all such $H$, the interference alignment equations are solvable.

\begin{theorem}
   For 4-user interference channel with reciprocal feedback channel ($G = H^T$), if $H$ is symmetric channel  and all diagonal entries are zero, i.e., $H$ can be written as
   \begin{align}
        H = \left(
  \begin{array}{cccc}
    0 & h_1 & h_2 & h_3  \\
    h_1 & 0 & h_4 & h_5 \\
    h_2 & h_4 & 0 & h_6 \\
    h_3 & h_5 & h_6 & 0 \\
  \end{array}
\right),
    \end{align}
  then for generic $\{h_1,h_2,h_3,h_4,h_5,h_6\}$, i.e., for all $\{h_1,h_2,h_3,h_4,h_5,h_6\}$ except a set of $\{h_1,h_2,h_3,h_4,h_5,h_6\}$ which satisfies a nontrivial polynomial equation, there exists $D_1, D_2$ and $D_3$ such that $(HD_2 + HD_3 G D_1H )$ is proportional to the corresponding row of $H$ except the diagonal entries.
 \end{theorem}

 \begin{IEEEproof}
   We derive the interference alignment equations and then use commercial computer algebra system Maple to compute the corresponding symbolic  Gr\"{o}bner basis, which does not contain a nonzero scalar. Therefore, due to Corollary \ref{cor:2} in Appendix \ref{app:1}, we conclude that for generic  $\{h_1,h_2,h_3,h_4,h_5,h_6\}$, there exists $D_1, D_2$ and $D_3$ such that $(HD_2 + HD_3 G D_1H )$ is proportional to the corresponding row of $H$ except the diagonal entries.
 \end{IEEEproof}

\subsection{$K = 5$ and $ K = 6$}

Recall that the number of variables $N_v$ is $3K$, the number of equations $N_e$ is $K(K-2)$ and the number of inequalities $N_{ie}$ is $K$. $N_e$ grows faster than $N_v$ as $K$ increases, and when $K \ge 5$, $N_v \le N_e$. We expect that when the number of variables is not more than the number of inequations, a valid solution may not exist. Indeed, we confirm this conjecture for $K = 5$ and $6$ via computing Gr\"{o}bner basis for random numeric $H$ and $G$ and finding that $1$ is in the Gr\"{o}bner basis in all instances.

\begin{claim}
  For $K$-user interference channel with reciprocal feedback channel ($G = H^T$) where $K=5$ or $6$, for generic matrix $H$, there does not exist $D_1,D_2$ and $D_3$ such that each row of $(HD_2 + HD_3 G D_1H )$ is proportional to the corresponding row of $H$ except the diagonal entries.
\end{claim}

\begin{claim}
  For $K$-user interference channel with out-of-band feedback feedback channel where $K=5$ or $6$, for generic matrices $H$ and $G$, there does not exist $D_1,D_2$ and $D_3$ such that each row of $(HD_2 + HD_3 G D_1H )$ is proportional to the corresponding row of $H$ except the diagonal entries.
\end{claim}

Since one reverse transmission from destinations to sources is not enough, in Phase 2 (interaction phase) we can let destinations make more than one reverse transmissions to sources. In one extreme case, if each destination sends what it receives in Phase 1 to all sources sequentially, where in total we use $K$ reverse transmissions\footnote{In fact, it is easy to see that $(K - 1)$ reverse transmissions are also sufficient.}, then each source can compute the  symbol signals sent out by all sources in the first phase, and then each source can do precoding over all symbols in the second phase, which reduces to a MIMO broadcast channel.

It turns out that when $K =5$ or $K = 6$, two reverse transmissions in Phase 2 is sufficient to do interference alignment while still preserving the desired signals. A natural three-phase scheme is as follows:
\begin{itemize}
  \item Phase 1: All sources send their independent symbols  simultaneously. And destinations get $y = H x + n$, where $n$ is the  additive noise of the channel.
  \item Phase 2-1: After receiving signals from sources in phase1, all destinations scale $y$ and send back to sources using the feedback channel. sources  get $f^1 = GD_1 y +  \text{noise}$.
  \item Phase 2-2: All destinations scale $y$ using a different coding matrix and send back to  sources. sources get $f^2 = GD_2 y +  \text{noise}$.
  \item Phase 3: Now each source has three sets of signals $x, f^1$ and $f^2$, and sources can send out a linear combination of $x, f^1$ and $f^2$ to destinations via the forward channel. destinations get
  \begin{align}
    y' &= H(D_3 x + D_4 f^1 + D_5 f^2) + \text{noise} \\
      & =  (HD_3 + HD_4 G D_1H + HD_5 G D_2H ) x + \text{noise}.
  \end{align}
\end{itemize}

With two reverse transmissions, the number of variables is increased to $5K$ from $3K$, and is more than $N_e = K(K-2)$ for $K = 5,6$. Extensive numeric simulations show that for generic matrix $H$ and $G$, there exists diagonal coding matrices $\{D_i\}_{1\le i \le 5}$ to align interferences while still preserving desired signals.

\begin{claim}
  For $K$-user interference channel with  reciprocal feedback channel ($G = H ^T$) where $K=5$ or $6$, for generic matrices $H$, there  exists $D_1,D_2, D_3, D_4$ and $D_5$ such that each row of $(HD_3 + HD_4 G D_1H + HD_5 G D_2H )$ is proportional to the corresponding row of $H$ except the diagonal entries.
\end{claim}

\begin{claim}
  For $K$-user interference channel with  out-of-band feedback channel where $K=5$ or $6$, for generic matrices $H$ and $G$, there  exists $D_1,D_2, D_3, D_4$ and $D_5$ such that each row of $(HD_3 + HD_4 G D_1H + HD_5 G D_2H )$ is proportional to the corresponding row of $H$ except the diagonal entries.
\end{claim}


\section{In-Band Interactive Interference Alignment}\label{sec:fullduplex}
In this section, we study how to exploit interaction in interference channel where all sources and destinations have full-duplex antennas so that all nodes can transmit and receive in the same band simultaneously. We call this model as in-band interactive alignment. This model captures the full range of possibilities of interaction, including source-cooperation, destination-cooperation and feedback, where these cooperation modes arise naturally in a fully-connected wireless network. We show a surprising result that the presence of these modes can highly simplify the nature of communication schemes. In particular, for the $3$-user interference channel with in-band interaction, these modes, the optimal sum degrees of freedom of $\frac{3}{2}$ can be achieved using the proposed alignment schemes with a simple two-phase scheme even when the channel coefficients are fixed. We also show a similar result is true even when each user has $M \geq 1$ antennas. 

	
\subsection{Interference Channels with $K=3,4$}
Recall in the system model of IFC, $H$ is the channel matrix from sources to destination,   $U$ is the channel matrix among sources, and  $W$ is the channel matrix among destinations. Consider the following simple two-phase transmission scheme.
\begin{itemize}
  \item Phase 1:  All sources send out signals $x$ simultaneously. After the transmission, sources get $f = Ux + \text{noise}$, and destinations get $ y = Hx + \text{noise}$.

  \item Phase 2: sources send out a linear combination of $x$ and $f$, and destinations send out  scaled version of $y$. Therefore, destinations get $H(D_1 x + D_2 f) + W D_3y + \text{noise} = (HD_1 + HD_2U + WD_3H) x + \text{noise}$, where $D_1,D_2$ and $D_3$ are diagonal coding matrices.
\end{itemize}

Note that in this transmission scheme no feedback channel from destinations to sources is required. Similarly, if each row of $(HD_1 + HD_2U + WD_3H)$  is proportional to the corresponding row of $H$ except all diagonal entries, then interferences at all destinations are aligned and all destinations can retrieve their designed signals from sources without any interference, and thus achieve the optimal $\frac{K}{2}$ degrees of freedom.

Our first result is that the above two-phase transmission scheme works for $K=3$ and $K=4$.

\begin{theorem}
	For $K =3$ and $K=4$, for generic channel matrices $U$, $H$ and $W$, there exists diagonal matrices $D_1,D_2$ and $D_3$ such that each row of $(HD_1 + HD_2U + WD_3H)$  is proportional to the corresponding row of $H$ except all diagonal entries.
\end{theorem}

\begin{IEEEproof}
	The interference alignment equations are linear in $D_1,D_2$ and $D_3$, so as in the proof of Theorem \ref{thm:3user} we can use dimension argument to show that there always exists a solution  to the system of linear interference alignment equations and it also satisfies the inequality constraints.
\end{IEEEproof}

Using the same dimension argument, we can prove that the above scheme does not work for $K = 5$ or bigger. 

\subsection{Multi-antenna IFC with  $K = 4$}

Next we show that for four-user  MIMO interference channel with in-band interaction can help design a simple transmission scheme to achieve the optimal $\frac{KM}{2}$ degrees of freedom.

For $K$-user MIMO interference channel where all sources and destinations are equipped with $M$ antennas,  \cite{CJ2008} shows that for $K=3, M>1$, vector space interference alignment can achieve the optimal $\frac{KM}{2} = \frac{3M}{2}$ degrees of freedom and channel diversity is not required.  \cite{BCT2011} and \cite{Luo11} prove that in general vector space MIMO interference alignment can at most get  $\frac{2KM}{K+1}$ degrees of freedom, which is strictly less than the optimal $\frac{KM}{2}$ when $K \ge 4$. We show that for four-user MIMO interference channel with in-band interaction, a simple two-phase transmission scheme can achieve the optimal $2M$ degrees of freedom at least  for all $M \le 15$.

Suppose source $s_i$ wants to send $M$ symbols $x_{(i-1)M+1}, x_{(i-1)M+2}, \dots, x_{iM}$ to destination $t_i$, for all $1 \le i \le 4$. Let
$x \triangleq (x_1,x_2,\dots,x_{4M})^T$. Consider a natural two-phase transmission scheme for 4-user MIMO IFC with in-band interaction, described as follows.
\begin{itemize}
  \item Phase 1:  All sources send out signals $x$ simultaneously, i.e., the $i$th antenna sends $x_i$ for all $1 \le i \le 4M$. After the transmission, sources get $f = Ux + \text{noise}$, and destinations get $ y^1 = Hx + \text{noise}$.

  \item Phase 2: sources do linear coding over $x$ and $f$, and send $D_1 x + D_2 f$, and destinations do linear coding over $y^1$ and send $D_3 y$, where $D_1, D_2$ and $D_3$ are block diagonal matrices with diagonal block size of $M \times M$. Therefore, destinations get
      \begin{align}
      y^2 &= H(D_1 x + D_2 f) + W D_3y^1 + \text{noise} \nonumber \\
       &= (HD_1 + HD_2U + WD_3H) x + \text{noise}.
       \end{align}
\end{itemize}

Note that since each node can only do coding over what signals it has or received from other nodes, the coding matrices $D_1, D_2$ and $D_3$ have to be block diagonal matrices with diagonal block size of $M \times M$.

A sufficient condition for all destinations to retrieve the desired symbols from the corresponding sources without interference is that at each antenna of  destination $t_i$, all interference  symbols are aligned except the desired symbols $x_{(i-1)M+1}, x_{(i-1)M+2}, \dots, x_{iM}$, for all $1 \le i \le 4$.

More precisely, let  $B \triangleq HD_1 + HD_2U + WD_3H $ and denote the index set $\{(i-1)M+1,(i-1)M+2,\dots,iM\}$ by $I_i$. Then for the destination  $t_i$, in Phase 1 it receives
\begin{align}
  y^1_m = \sum_{m'\in I_i}H_{mm'} x_m' + \sum_{j \notin I_i} H_{mj} x_j + \text{noise term}, \label{eqn:mimo1}
\end{align}
and in Phase 2 it receives
\begin{align}
  y^2_m = \sum_{m'\in I_i}B_{mm'} x_m' + \sum_{j \notin I_i} B_{mj} x_j + \text{noise term}. \label{eqn:mimo2}
\end{align}

The first term on the RHS of \eqref{eqn:mimo1} and  \eqref{eqn:mimo1} correspond to the desired signals, and the second term are interferences. If at each antenna of $t_i$, all interferences are aligned, i.e.,
for all $m \in I_i$, there exists $\lambda_m \in \C$,
such that
\begin{align}
  \frac{B_{mj}}{H_{mj}} = \lambda_m, \quad \forall j \notin I_i, \label{eqn:MIMOeqn1}
\end{align}
then by computing $\tilde{y}_m \triangleq y^2_m - \lambda_m y^1_m$, $t_i$ can cancel all interferences
\begin{align}
  \tilde{y}_m &\triangleq y^2_m - \lambda_m y^1_m \\
  &= \sum_{m'\in I_i}(B_{mm'} -  \lambda_m B_{mm'})x_m'  + \text{noise term}.
\end{align}

Now $t_i$ gets $k$ linear equations in $k$ variables $\{x_m\}_{m \in I_i}$. If the coefficient matrix of the $k$ linear equations has full rank, then $t_i$ can retrieve all desired signals $\{x_m\}_{m \in I_i}$ corrupted by noise only.

The condition that the coefficient matrix has full rank is equivalent to the condition that the determinant of the coefficient matrix is nonzero. Therefore, to make the transmission scheme work, we need to find block diagonal coding matrices $D_1,D_2,D_3$ to satisfy the interference alignment linear equations \eqref{eqn:MIMOeqn1} and make the determinants of coefficient matrices be nonzero. Without introducing the auxiliary variables $\lambda_m$, we have $KM((K-1)M-1)$ interference alignment equations, which are linear in terms of $D_1, D_2$ and $D_3$. Since the equations are linear, there exists analytical solution to $D_1, D_2$ and $D_3$ which are rational functions of $U, H$ and $W$. Then, we can plug the analytical solution into the expression of the determinant of the coefficient matrix and check whether it is zero. We have checked that for $K = 4$, the determinants of the coefficient matrices are  nonzero expressions for all $M \le 15$.
\begin{theorem}\label{thm:MIMO}
  For 4-user MIMO interference channel where each node has $M$ antennas, then for generic matrices $H, U$ and $W$, the two-phase scheme can achieve the optimal $\frac{K}{2}$ degrees of freedom, at least for $M \le 15$.
\end{theorem}

Note that for two-phase scheme, in total we have $N_v = 3KM^2$ variables and $N_e = KM((K-1)M-1)$ linear equations. When $K = 4$, $N_v = 12M^2 > N_e = 12M^2 - 4M$ for all $M$, which means we have more variables than equations, and thus we believe Theorem \ref{thm:MIMO} holds for any $M$.

\section{Numeric Finite SNR Performance Evaluation}\label{sec:simulation}   

In this section, we present simulation results on the finite SNR performance of our three-phase interactive interference alignment scheme with out-of-band reverse (feedback) transmissions  for $3$-user and $4$-user interference channels, where the channel parameters are drawn from i.i.d. complex Gaussian distribution. We assume both sources and destinations have the same power constraint. We formulate the problem as a nonconvex optimization problem, and use generic local algorithm to maximize the sum rates among all possible interference alignment solutions while satisfying power constraints. For each SNR, we simulate over 50 channel realizations and compute the average of maximum sum rates. We compare the maximum rates achieved via interactive interference alignment with the standard time sharing scheme.

\begin{figure}[t]
\centering
\includegraphics[scale=0.135]{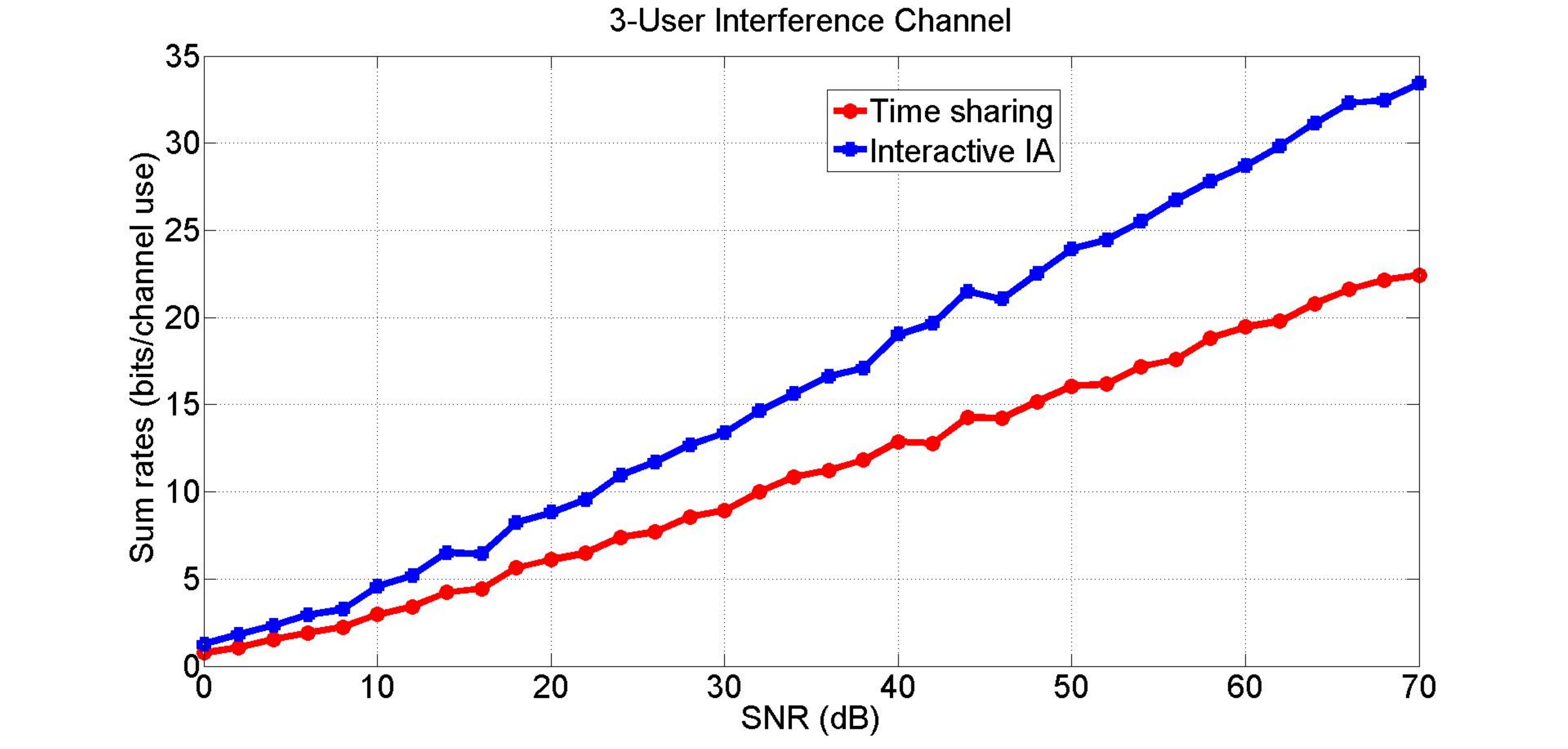}
\caption{Numeric Finite SNR Performance Simulations for 3-User Interference Channel with Out-of-Band Reverse (Feedback) Channel.}
\label{fig:sim1}
\end{figure}

\begin{figure}[t]
\centering
\includegraphics[scale=0.135]{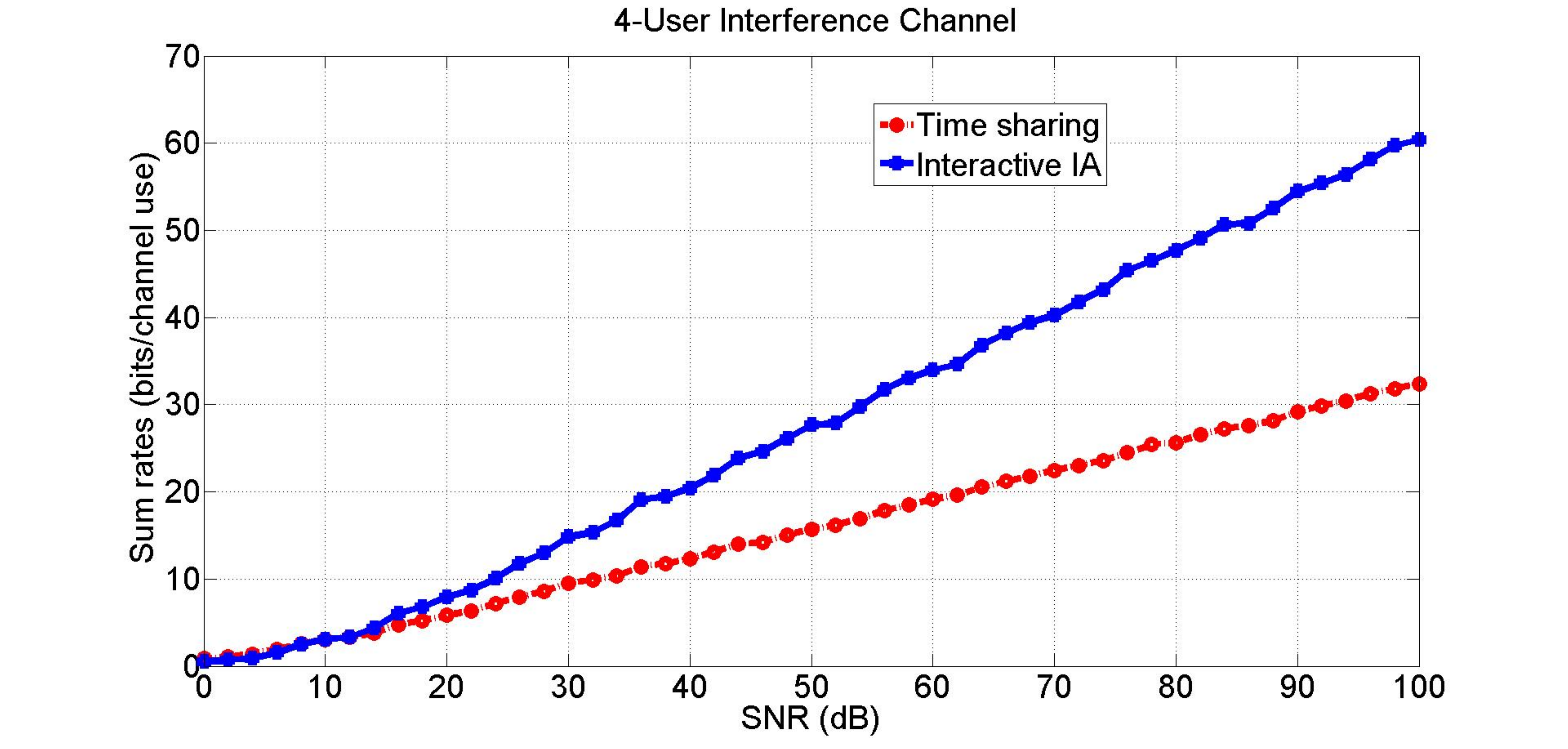}
\caption{Numeric Finite SNR Performance Simulations for 4-User Interference Channel with Out-of-Band Reverse (Feedback) Channel.}
\label{fig:sim2}
\end{figure}

For $K=3$ and $K = 4$, our interactive interference alignment scheme can achieve the optimal $\frac{K}{2}$ degrees of freedom, while time sharing scheme only gets $1$ degree of freedom. Therefore, at high SNR, we expect our scheme has  $50$ percent improvement for $K = 3$ and $100$ percent improvement for $K=4$ over time sharing scheme, and this matches our simulation results in Figure \ref{fig:sim1} and Figure \ref{fig:sim2}.

\section{Discussions and Open Problems}\label{sec:discussion}

In this section, we give a general multi-phase interactive communication scheme for  $K$-user interference channel with large $K$ and discuss how $O(\sqrt{K})$ degrees of freedom might be achieved with  reverse transmissions taken into account. We also discuss some open problems following this line of work.

\subsection{Multi-Phase Transmission Scheme for General $K$}\label{sec:generalcase}

In Section \ref{sec:scheme}, we have shown that for general $K$ we can achieve the optimal $\frac{K}{2}$ degrees of freedom using the three-phase scheme with $(K-1)$ reverse transmissions in Phase 2. However, if we take reverse transmissions into account, the total number of transmissions is $2 + (K-1) = K+1$, and thus the total degrees of freedom is $\frac{K}{K+1} < 1$. In this section, we give a multi-phase transmission scheme and use equation-variable counting argument to show that we may potentially achieve $O(\sqrt{K})$ degrees of freedom even if we take reverse transmissions into account.

\textbf{Multi-Phase Transmission Scheme}
\begin{itemize}
	\item Forward Phase 1: All TXs send their independent symbols  simultaneously. And RXs get $y^1 = H x + \text{noise}$.

	\item Reverse Phase 1: All RXs scale $y$ and send back to TXs using the feedback channel.  TXs get $f^1 = GD_1 y + \text{noise}$.

	\item Forward Phase 2: All TXs send a linear combination of $x$ and $f^1$ to RXs. And RXs get $y^2 = H (D_2x + D_3 f^1) + \text{noise}$.

	\item Reverse Phase 2: All RXs send a linear combination of $y^1$ and $y^2$ to RXs.   TXs get $f^2 = G(D_4 y^1 + D_5 y^2) + \text{noise}$.

	\item \dots

	\item Forward Phase i: All TXs send a linear combination of $x, f^1, f^2, \dots, f^{i-1}$ to RXs, using $i$ diagonal coding matrices, and RXs get $y^i$.

	\item Reverse Phase i: All RXs send a linear combination of $y^1, y^2, \dots, y^{i}$ to RXs, using $i$ diagonal coding matrices, and RXs get $y^i$.

	\item \dots

	\item Forward Phase N: All TXs send a linear combination of $x, f^1, f^2, \dots, f^{N-1}$ to RXs, using $N$ diagonal coding matrices, and RXs get $y^N$.

\end{itemize}

The number of coding matrices is
\begin{align}
	& 1 + 2 + 2 + 3 + 3 + \cdots + (N-1) + (N-1) + N  \nonumber \\
=& \ N^2 - 1,
\end{align}
so the total number of variables $N_v$ is $(N^2-1) K$. On the other hand, the total number of interference alignment equations $N_e$ is $K(K-2)$. If we want to have $N_v \ge N_e$, then $N \ge \sqrt{K-1}$, and thus $N = \Omega(\sqrt{K})$. If there exists a solution to the corresponding system of polynomial equations, then the sum degrees of freedom achievable is
\begin{align}
	\frac{K}{2N} \le \frac{K}{2\sqrt{K-1}} = O(\sqrt{K}).
\end{align}

Therefore, by equation-variable counting, we conjecture that the multi-phase interactive interference alignment may achieve $O(\sqrt{K})$ degrees of freedom for general $K$.

\subsection{Open Problems}\label{sec:openprob}

We list several interesting open problems following this line of work on interactive interference alignment.

\begin{itemize}
	\item \textbf{Constructive Proof.} While we have used the generic Gr\"{o}bner basis tool to show whether the interference alignment equations are solvable or not, it will be more interesting to derive an analytical solution, like what we have done in the case where $H$ can be represented as the sum of a diagonal matrix and a rank $1$ matrix.

	\item \textbf{Efficient Algorithm and Optimization.} In theory we can use the Buchberger's algorithm to get an interference alignment solution, but the computational complexity is high. So one natural question is whether we can design a much more efficient algorithm to solve the interference alignment equations. In addition, as presented in Section \ref{sec:simulation}, to maximize the sum rates using interactive interference alignment, we formulate a nonconvex optimization problem and use generic local algorithm to solve it. Can we design an efficient algorithm to solve the optimization problem so that it can be used in practice (in a distributed way)?

	\item \textbf{Out-of-Band interactive IA for general $K$.} We give an equation-variable counting argument to show why $O(\sqrt{K})$ degrees of freedom might be achievable using interactive interactive alignment with reverse transmissions into account. Can one rigorously prove or disprove it? Can we design a new transmissions scheme to get degrees of freedom linearly growing with $K$ using the same idea of interactive interference alignment?

	\item \textbf{In-Band interactive IA for general $K$.} We have shown for small $K$, e.g., $K=3$ and $4$, interactive IA with full-duplex radios can achieve the optimal degrees of freedom, while for general $K$ the current two phase scheme does not work for $K \ge 5$. It is interesting to study how to extend the two phase interactive IA to general $K$-user interference channel with in-band interaction. In particular, the two-phase interactive scheme did not take advantage of the presence of the feedback channels from the source to the destination. How to use these feedback channels in order to achieve better performance for $K \geq 5$ is an open question. In general it may be needed to do a scheme with many phases; it is interesting to understand how to analyze such schemes. 

	\item \textbf{Interactive IA with relays.} Relays naturally fit into our model, for both half-duplex and full-duplex radios. Through interaction among relays, sources and destinations, relays can help sources do interference alignment. For instance, for half-duplex $K$-user interference channel, with $aK^2$ relays, for some constant $a$, in the natural three phase scheme with one reverse transmission from the destinations, relays can help align all interferences at each destination. Can one characterize the minimum number of relays needed, or design a new transmission scheme that achieves alignment using $o(K^2)$ relays?

\item \textbf{In-band Interaction with Bi-directional Traffic.} When we studied in-band interaction, we assumed that even though the channels permit bi-directional communication, the traffic model was uni-directional. Indeed, there are some practical scenarios, where the channel is bidirectional, but the traffic is unidirectional at a given point of time. For example, it is well known that downloads are higher than uploads from a base station and therefore we can think of a scenario, where the three transmitters are base stations and the three receivers are mobile users. In this case, our proposed alignment scheme may be useful. However, the general case of in-band interaction in the presence of bi-directional traffic is interesting to study, and whether there are intelligent schemes that can simultaneously achieve $\frac{K}{2}$ degrees of freedom on both directions is an open question. 

\end{itemize}

\section{Conclusion}\label{sec:conclusion}

We study new channel models where interaction among sources and destinations is enabled, e.g., both sources and destinations can talk to each other.  The interaction can come in two ways: 1) for half-duplex radios, destinations can talk back to sources using simultaneous  out-of-band transmission or in band half-duplex transmission; 2) for full-duplex radios, both sources and destinations can transmit and listen in the same channel simultaneously.
 Although \cite{CJ2009} shows that for interference channel, relays, feedback, and full-duplex operation cannot improve the degrees of freedom beyond $\frac{K}{2}$,   we demonstrate that the interactions among sources and destinations enables flexibility in designing simple interference alignment scheme and in several cases achieves the optimal degrees of freedom.

 We present a three-phase interactive communication scheme for half-duplex $K$-user interference channel with small $K$, a two-phase interference communication scheme for full-duplex	$K$-user interference channel with small $K$, and show that the interactive interactive alignment scheme can achieve the optimal degrees of freedom for several classes of IFC, including half-duplex $3$-user  IFC, full-duplex  $3$-user and $4$-user IFC, and full-duplex $4$-user MIMO IFC. We do extensive numeric simulations and show the proposed interactive interactive alignment scheme also works for some other classes of IFC empirically. We use tools from algebraic geometry to show why success of numeric simulations can suggest the scheme should work well for almost all channel parameters in a rigorous way.  Simulation results on the finite SNR performance of interactive interactive alignment for $3$-user and $4$-user half-duplex interference channels are presented. Finally, we discuss a general multi-phase interactive communication scheme for  $K$-user interference channel with large $K$, and list several open problems.

\appendices

\section{Proof of Theorem \ref{thm:special}}\label{app:special}
\begin{IEEEproof}[Proof of Theorem \ref{thm:special}]
Following the notation introduced in Section \ref{sec:problemformulation}, equivalently, we need to prove that there exists diagonal coding matrices $D_1, D_2$ and $D_3$ such that $(HD_2 + HD_3 G D_1H )$ is a diagonal matrix and each diagonal entry is strictly nonzero.

We give a constructive proof. First, we expand $(HD_2 + HD_3 G D_1H )$ and get
	\begin{align}
		& HD_2 + HD_3 G D_1H  \\
		=& (D+uv^T)D_2 + (D+uv^T)D_3 (D+vu^T)D_1(D+uv^T) \\
	 =& DD_2+ uv^T D_2 + \nonumber \\
        & (DD_3 + uv^TD_3)(DD_1 + vu^TD_1)(D+uv^T) \\
	 =& DD_2+ uv^TD_2 + (DD_3 D D_1 + uv^T D_3 D D_1 + \nonumber \\
         & uv^T D_3 vu^TD_1 + DD_3vu^TD_1) (D+uv^T)\\
       =& DD_2 + uv^TD_2 + DD_3 D D_1D +  uv^T D_3 D D_1 D + \nonumber \\ 
       & uv^T D_3 vu^TD_1D + DD_3vu^TD_1D +  \nonumber \\ 
       & DD_3 D D_1uv^T + uv^T D_3 D D_1 uv^T +  \nonumber \\ 
       & uv^TD_3 vu^TD_1uv^T + DD_3vu^TD_1uv^T.
\end{align}

Define $U \triangleq \text{diag}(u)$, i.e., $U$ is a diagonal matrix where the $i$th diagonal entry is the $i$th component of $u$. Similarly, define $V \triangleq \text{diag}(v)$. Set
\begin{align}
D_1 &= D^{-1} U^{-1}V,\\
D_3 &= D^{-1} V^{-1}U.
\end{align}
Then
\begin{align}
D_3 D D_1 D &= I, \\
u^TD_1 D &= v^T,\\
DD_3 v &= u,
\end{align}
where $I$ is the identity matrix.

Therefore,
\begin{align}
	& HD_2 + HD_3 G D_1H  \\
       =& DD_2 + uv^TD_2 + D +  uv^T  +  uv^T D_3 v v^T \nonumber \\ 
       & + u v^T + uv^T + u u^T D_1 uv^T +  \nonumber \\ 
        &   uv^TD_3 vu^TD_1uv^T + uu^TD_1uv^T \\
       =& D(D_2+I) + uv^TD_2 + (3+ 2v^TD_3 v+ u^TD_1u + \nonumber \\ 
       & v^TD_3v u^TD_1u  )uv^T \\
       =& D(D_2+I) + uv^TD_2 + \nonumber \\ 
       & (3+ 3v^TD^{-1}u + (v^TD^{-1}u )^2) uv^T.
\end{align}

Define $\alpha = 3+ 3v^TD^{-1}u + (v^TD^{-1}u )^2$, and set
\begin{align}
	D_2 = - \alpha I.
\end{align}
Then
\begin{align}
	 HD_2 + HD_3 G D_1H  = (1-\alpha) D.
\end{align}
Since $\alpha \neq 1$, $(HD_2 + HD_3 G D_1H)$  is a diagonal matrix and each diagonal entry is nonzero.

\end{IEEEproof}

\section{Mathematical background on algebraic geometry}\label{app:1}

 In the section we present some  basic terminologies and results in algebraic geometry. In particular, we introduce the Gr\"{o}bner basis  and Buchberger's algorithm, which are standard tools to check whether a system of polynomial equations have solutions or not. For more details on algebraic geometry, we refer the reader to the excellent textbook \cite{AGbook}.

Recall that we use $k$ to denote a field,  $k[\xi_1,\xi_2,\dots,\xi_n]$ to denote the set of all polynomials in $\xi_1,\xi_2,\dots,\xi_n$ with coefficients in $k$, and use $V(f_1,\dots,f_s)$ to denote the affine variety defined by polynomials $f_1,\dots,f_s \in k[\xi_1,\xi_2,\dots,\xi_n]$.

\begin{definition}[Definition 2 on page 30 of \cite{AGbook}]
  Let $f_1,\dots,f_s$ be polynomials in $k[\xi_1,\xi_2,\dots,\xi_n]$. Then we denote the ideal generated by $f_1,\dots,f_s$ as
  \begin{align}
    \left \langle  f_1,\dots,f_s \right \rangle  \triangleq \{ \sum_{i=1}^s h_i f_i\ | \  h_1,\dots,h_s \in k[\xi_1,\xi_2,\dots,\xi_n] \}.
  \end{align}
\end{definition}

It is straightforward to see that
\begin{align}
V(f_1,\dots,f_s) = V(\left \langle  f_1,\dots,f_s \right \rangle ).
\end{align}

\begin{theorem}[ The Weak Nullstellensatz on page 170 of \cite{AGbook}]  \label{thm:nullstellensatz}
  Let $k$ be an algebraically closed field and let $I \subset k[\xi_1,\xi_2,\dots,\xi_n]$ be an ideal satisfying $V(I) = \emptyset$. Then $I = k[\xi_1,\dots,\xi_n]$.
\end{theorem}

\begin{corollary}\label{cor:1}
  Let $f_1,\dots,f_s$ be polynomials in $k[\xi_1,\xi_2,\dots,\xi_n]$, where $k$ is an algebraically closed field. Then $V(f_1,\dots,f_s) = \emptyset$ if and only if $1 \in \left \langle  f_1,\dots,f_s \right \rangle $.
\end{corollary}
\begin{IEEEproof}
  If $1 \in \left \langle  f_1,\dots,f_s \right \rangle $, then $\left \langle  f_1,\dots,f_s \right \rangle  =  k[\xi_1,\dots,\xi_n]$, and thus $V(f_1,\dots,f_s) = V(\left \langle  f_1,\dots,f_s \right \rangle ) = V(k[\xi_1,\dots,\xi_n]) = \emptyset$.

  On the other hand, if $V(f_1,\dots,f_s) = \emptyset$, equivalently, $V(\left \langle  f_1,\dots,f_s \right \rangle ) = \emptyset$, then by Theorem \ref{thm:nullstellensatz} we have
  \begin{align}
    \left \langle  f_1,\dots,f_s \right \rangle  = k[\xi_1,\dots,\xi_n],
  \end{align}
  and thus $1 \in \left \langle  f_1,\dots,f_s \right \rangle $.

\end{IEEEproof}

Therefore, given polynomials $f_1,\dots,f_s \in k[\xi_1,\dots,\xi_n]$, where $k$ is an algebraically closed field, to check whether $V(f_1,\dots,f_s)$ is nonempty, i.e., whether there exists solution  to the system of polynomial equations
\begin{align}
  f_i = 0, \forall 1 \le i \le s,
\end{align}
 we only need to check whether $1 \in \left \langle  f_1,\dots,f_s \right \rangle $. This is the ideal membership problem, which is well studied in algebraic geometry. The basic tool we will use is \emph{Gr\"{o}bner basis} \cite{Buchberger}. Below we give a short introduction to \emph{Gr\"{o}bner basis} and show how to use \emph{Gr\"{o}bner basis} to solve the ideal membership problem.

 \begin{definition}[Definition 7 on page 59 of \cite{AGbook}]
   Let $f = \sum_{\alpha} a_{\alpha}x^{\alpha}$ be a nonzero polynomial in $k[\xi_1,\dots,\xi_n]$ and let $>$ be a monomial order. Then we define the \textbf{multidegree} of $f$ as
   \begin{align}
     multideg(f) \triangleq \max (\alpha: a_{\alpha} \neq 0 ),
   \end{align}
   where the maximum is taken with respect to the monomial order $>$.

   And we define the \textbf{leading term} of $f$ as
   \begin{align}
      LT(f) \triangleq a_{multideg(f)} \xi^{multideg(f)}.
   \end{align}

 \end{definition}

Given an ideal $I$, we use $LT(I)$ to denote the set of leading terms of elements of $I$, i.e.,
\begin{align}
  LT(I) \triangleq \{LT(f) |  f \in I  \}.
\end{align}

Due to Theorem \ref{thm:hilbert}, every ideal in $k[\xi_1,\dots,\xi_n]$ has a finite generating set.
\begin{theorem}[Hilbert Basis Theorem on page 76 of \cite{AGbook}] \label{thm:hilbert}
  Every ideal $I \subset k[\xi_1,\dots,\xi_n]$ has a finite generating set. That is, $I = \left \langle g_1,\dots,g_t \right \rangle$ for some $g_1,\dots,g_t \in I$.
\end{theorem}

\begin{definition}[Definition 5 on page 77 of \cite{AGbook}]    
  Fix a monomial order. A finite subset $G = \{g_1,\dots,g_t\}$ of an ideal $I$ is said to be a \textbf{Gr\"{o}bner basis} if
  \begin{align}
    \left \langle LT(g_1), \dots, LT(g_t)\right \rangle = \left \langle LT(I)\right \rangle.
  \end{align}
\end{definition}

Equivalently, a set $\{g_1,\dots,g_t\} \subset I$  is a Gr\"{o}bner basis of $I$ if and only if the leading term of any element of $I$ is divisible by one of the $LT(g_i)$ \cite{AGbook}.

\begin{proposition}[Proposition 1 on page 82 of \cite{AGbook}]
  Let $G = \{g_1,\dots,g_t\}$ be a Gr\"{o}bner basis for an ideal $I \subset k[\xi_1,\dots,\xi_n]$ and let $f \in k[\xi_1,\dots,\xi_n]$. Then there is a unique $r \in k[\xi_1,\dots,\xi_n]$ with the following two properties:

  (i) No term of $r$ is divisible by any of $LT(g_1), \dots, LT(g_t)$.

  (ii) There is $g \in I$  such that $f = g + r$.

  In particular, $r$ is the remainder on division of $f$ by $G$ no matter how the elements of $G$ are listed when using the division algorithm.
\end{proposition}

An important property of  Gr\"{o}bner basis is that given a polynomial $f$, we can check whether $f$ is in the corresponding ideal via division of $f$ by $G$. More precisely,

\begin{corollary}[Corollary 2 on page 82 of \cite{AGbook}]\label{cor:3}
  Let $G = \{g_1,\dots,g_t\}$ be a  Gr\"{o}bner basis for an ideal $I \subset k[\xi_1,\dots,\xi_n]$ and let $f \in k[\xi_1,\dots,\xi_n]$. Then $f \in I$ if and only if the remainder on division of $f$ by $G$ is zero.
\end{corollary}

\begin{corollary}\label{cor:2}
   Let $G = \{g_1,\dots,g_t\}$ be a Gr\"{o}bner basis for an ideal $I \subset k[\xi_1,\dots,\xi_n]$. Then $1 \in I$ if and only if
   \begin{align}
     g_i = \lambda,
   \end{align}
   for some $i \le t$ and nonzero $\lambda \in k$.
\end{corollary}

\begin{IEEEproof}
  We first prove the easier direction. If for some $i, g_i = \lambda \neq 0$, then $\left \langle g_i \right \rangle  =  k[\xi_1,\dots,\xi_n]$. So $I = k[\xi_1,\dots,\xi_n]$ and thus $1 \in I$.

  For the other direction, suppose that  $1 \in I$, but for each $i \le t$, the leading term of $g_i$ has a nonzero order, i.e., $g_i$ does not have the form $g_i = \lambda$ for some nonzero $\lambda \in k$. Then after we do division of $1$ by $G$, the remainder will be $1$, so the remainder is nonzero, and thus $1 \notin I$ by Corollary \ref{cor:3}, which leads to a contradiction. Therefore, for some $i$, $g_i$ can be written as  $g_i = \lambda$  for some nonzero $\lambda \in k$.
\end{IEEEproof}

Due to Corollary \ref{cor:1} and Corollary \ref{cor:2}, to check whether there exists solution to a system of polynomial equations over algebraically closed field, we only need to compute a Gr\"{o}bner basis of the corresponding ideal and check whether a nonzero scalar is in the Gr\"{o}bner basis.

The standard approach to computing a Gr\"{o}bner basis is Buchberger's algorithm  \cite{Buchberger}, which can be viewed as a generalization of  Gaussian elimination for linear systems.

\begin{theorem}[Buchberger's Algorithm on page 90 of \cite{AGbook}]
Let $I=\left \langle f_1,f_2,\dots,f_s \right \rangle \neq \{0\}$ be a polynomial ideal. Then a Gr\"{o}bner basis for $I$ can be constructed in a finite number of steps by the following algorithm:

\begin{algorithm}
\caption{Buchberger's algorithm to compute Gr\"{o}bner basis }
\begin{algorithmic}
\State Input: $F = (f_1,f_2,\dots,f_s) $
\State Output: a Gr\"{o}bner basis $G = (g_1,g_2,\dots,g_t)$ for $I$, with $F \subset G$
\\
\State Initialization: $G \gets F$

\Repeat
\State $G' \gets G$
\For{each pair $\{p,q\}, p \neq q$ in $G'$}
\State $\alpha \gets \text{multideg}(p)$, and $ \beta \gets \text{multideg}(q))$
\State $\gamma \gets (\gamma_1,\dots,\gamma_n)$, where $\gamma_i = max(\alpha_i,\beta_i)$ for each $i$.
\State $S(p,q) \gets \frac{\xi^\gamma}{LT(p)}p - \frac{\xi^\gamma}{LT(q)}q$ (note that leading terms will be canceled)
\State $S \gets {\overline{S(p,q)}}^{G'}$, where ${\overline{S(p,q)}}^{G'}$ the remainder on division of   $S(p,q)$ by $G'$ using multivariate division algorithm
\If{$S \neq 0$}
  \State $G \gets G \cup {S}$
\EndIf
\EndFor

\Until{G = G'}
\end{algorithmic}
\end{algorithm}

\end{theorem}

Buchberger's algorithm can be proven to terminate in finite number of steps, and one important consequence is that if the coefficients in the polynomial equations are rational functions of variables $\{h_i\}$, then  except a set of $\{h_i\}$ which satisfies a nontrivial polynomial equations on $\{h_i\}$ and thus has a measure 0,  either for all $\{h_i\}$ the polynomial equations have a solution, or for all $\{h_i\}$ the polynomial equations do not have a solution. This result is clearly stated in Theorem  \ref{thm:equivalsym_num} in Section \ref{sec:math}.

\section{Proof of Theorem \ref{thm:equivalsym_num}}\label{app:proof}

\begin{IEEEproof}[Proof of Theorem \ref{thm:equivalsym_num}]
Suppose all coefficients in $f_1,\dots,f_s$ are in symbolic form, and we run Buchberger's algorithm to compute  $G = \{g_1,\dots,g_t\}$, the symbolic Gr\"{o}bner basis for  $\left \langle f_1,f_2,\dots,f_s \right \rangle$. In each step of Buchberger's algorithm, we will get new derived polynomials with symbolic coefficients, which are also rational functions of $h_1,h_2,\dots,h_m$. Since Buchberger's algorithm is proven to terminate in finite number of steps, the set of all symbolic coefficients, which are  all rational functions of $h_1,h_2,\dots,h_m$, is finite. Let $e$ be the product of all numerators and denominators of the set of symbolic coefficients, and thus $e$ is a nontrivial polynomial of $h_1,h_2,\dots,h_m$.

For all numeric $ (h_1, h_2, \dots, h_m) \in \C^{1 \times m} $ such that
$e(h_1, h_2, \dots, h_m) \neq 0$, the structure of numeric Gr\"{o}bner basis is the same as the structure of symbolic Gr\"{o}bner basis, and thus either $V(f_1,\dots,f_s) = \emptyset$ or $V(f_1,\dots,f_s) \neq \emptyset$, which depends on whether the symbolic Gr\"{o}bner basis contains a nonzero constant.

 Therefore, except the set of $\{(h_1,h_2,\dots,h_m) \  |\  e(h_1,h_2,\dots,h_m) = 0 \}$, which has a measure 0, either for all $ (h_1, h_2, \dots, h_m) \in \C^{1 \times m} $, $V(f_1,\dots,f_s) \neq \emptyset$, or for all $ (h_1, h_2, \dots, h_m) \in \C^{1 \times m} $, $V(f_1,\dots,f_s) = \emptyset$.
\end{IEEEproof}

\section*{Acknowledgment}\label{sec:acknowledgment}
We gratefully acknowledge funding from the National Science Foundation via grant CCF 1017430 and grant  ECCS-1232257, and Intel research grant.

\bibliographystyle{IEEEtran}
\bibliography{reference}

\end{document}